\documentclass[11pt,a4j]{article}

\usepackage[dvipdfmx]{graphicx}
\usepackage{color}
\usepackage{cite}
\makeatletter
\def\slash#1{{\mathpalette\c@ncel{#1}}} 
\makeatother

\newcommand\beq{\begin{eqnarray}}
\newcommand\eeq{\end{eqnarray}}

\newcommand\la{\langle}
\newcommand\ra{\rangle}

\def\Pslash{\rlap/{\mkern-1mu P}}
\def\yslash{\rlap/{\mkern-1mu y}}

\def\nslash{\slash{\mkern-1mu n}}

\def\Nhat{\widehat{N}}
\def\Dhat{\widehat{D}}

\def\Hhat{\widehat{H}}

\def\Ghat{\widehat{G}}
\def\DGhat{\Delta\widehat{G}}
\def\Hhat{\widehat{H}}
\def\DHhat{\Delta\widehat{H}}
\def\3Tb{3\bar{T}}

\def\Dtilde{\widetilde{D}}

\begin{document}

\begin{flushright}
\today
\end{flushright}
\vspace*{15mm}
\begin{center}
{\LARGE \bf Exact Relations for Twist-3 Gluon Distribution and\\[10pt] 
Fragmentation Functions from Operator Identities}
\vspace{1.5cm}\\
 {\sc Yuji Koike$^2$, Kenta Yabe$^3$ and Shinsuke Yoshida$^1$}
\\[0.7cm]
\vspace*{0.1cm}
{\it $^1$ Institute of Quantum Matter and School of Physics and Telecommunication Engineering, 
South China Normal University, Guangzhou 510006, China}

\vspace{0.2cm}

{\it $^2$ Department of Physics, Niigata University, Ikarashi, Niigata 950-2181, Japan}

\vspace{0.2cm}

{\it $^3$ Graduate School of Science and Technology, Niigata University,
Ikarashi, Niigata 950-2181, Japan}
\\[3cm]

{\large \bf Abstract} \end{center}
\vspace{0.2cm}

\noindent
We perform a systematic study on the twist-3 gluon distribution and fragmentation functions
which appear in the collinear twist-3 factorization for hard inclusive processes. 
Three types of twist-3 distribution and fragmentation functions, i.e., intrinsic, 
kinematical and dynamical ones, which
are necessary to describe all kinds of twist-3 cross sections, are related
to each other 
by the operator identities based on the QCD equation of motion and the Lorentz invariance properties
of the correlation functions.  
We derive the exact relations
for all twist-3 gluonic distribution and fragmentation functions for a spin-1/2 hadron. 
Those relations allow one to
express intrinsic and kinematical twist-3 gluon functions in terms of the twist-2 and dynamical twist-3
functions, which provides a basis for the renormalization of intrinsic and kinematical twist-3
functions. 
In addition, those model independent relations are crucial to guarantee gauge invariance and 
frame independence properties of the twist-3 cross sections.


\newpage
\section{Introduction}

During the past few decades twist-3 effects in (semi-)inclusive processes have been receiving great attention, 
in that they show up as a leading contribution to interesting spin asymmetries and reveal an 
important role of
multi-parton correlations in the scattering processes which shed new lights on the hadron structure.  
By now theoretical methods for dealing with those twist-3 effects have been developed
and widely used to derive many relevant twist-3 cross section formula.  
Such theoretical studies include those for $g_2$-structure function of the nucleon measured in
deep-inelastic scattering\,\cite{Jaffe:1989xx,Belitsky:2000pb},   
single spin asymmetries (SSA)
for a hadron or (virtual) photon 
production in proton-proton (nucleus) collisions with one proton transversely polarized, 
$p^\uparrow p\to h X$ ($h=\pi,\, D,\, \gamma,\,\gamma^*$ etc)\,
\cite{Qiu:1991wg,Qiu:1998ia,Kanazawa:2000hz,Ji:2006vf,Kouvaris:2006zy,Koike:2007rq,Koike:2009ge,
Vogelsang:2009pj,
Koike:2011b,Koike:2011nx,Kang:2010zzb,Kanazawa:2011er,Metz:2012ct,Beppu:2013uda,Kanazawa:2014nea}, 
$p^\uparrow A\to hX$\,\cite{Kang:2011ni,Hatta:2016wjz,Hatta:2016khv,Benic:2018moa,Benic:2018amn}, 
and semi-inclusive deep-inelastic scattering (SIDIS), $ep^\uparrow\to ehX$\,
\cite{Ji:2006br,Eguchi:2006qz,Eguchi:2006mc,Yuan:2009dw,Beppu:2010qn,Koike:2011a,
Kanazawa:2013uia,Yoshida:2016tfh,Xing:2019ovj,Benic:2019zvg},
SSA
in transversely polarized hyperon production in the unpolarized
proton-proton collision, $pp\to \Lambda^\uparrow X$\,\cite{Kanazawa:2001a,Zhou:2008,
Koike:2015zya,Koike:2017fxr,Yabe:2019awq,Kenta:2019bxd} and in $e^+e^-$-collision, 
$e^+e^-\to\Lambda^\uparrow X$\,\cite{Gamberg:2018fwy}, 
and the longitudinal-transverse double spin asymmetry $A_{LT}$
in the proton-proton collision,
$\vec{p}p^\uparrow\to \{h,\gamma^*\}\,X$\,
\cite{Jaffe:1991ra,Liang:2012rb,Hatta:2013wsa,Koike:2015yza,Koike:2016ura}, etc.  
Collinear twist-3 
parton distribution functions (DFs) and fragmentation functions (FFs)
which appear in these twist-3 factorization formula for the
cross sections 
no longer have probability interpretation
but represent multi-parton (quark-gluon or purely gluonic) correlations
in the hadrons or in the fragmentation processes.
The leading order (LO) 
evolution equations for the relevant
twist-3 functions have been also 
derived\cite{Balitsky:1987bk,Ali:1991em,Koike:1994st,Kodaira:1996md,Balitsky:1996uh,
Koike:1996bs,Belitsky:1997zw,Belitsky:1997by,Kang:2008ey,Zhou:2008mz,
Vogelsang:2009pj,Braun:2009mi,
Kang:2010xv,Ma:2011nd,Zhou:2015lxa,Yoshida:2016tfh}.

Collinear twist-3 DFs and FFs can be in general classified into three types:  
intrinsic, kinematical and dynamical ones\cite{Kanazawa:2015ajw} .  Although they all appear in
the calculation of the twist-3 cross section formula,
they are not independent from 
each other, but are related by QCD equation of motion (e.o.m.) and Lorentz invariance
properties of the correlation functions.  
One of the present authors (Y.K.) 
performed a systematic study on the twist-3 quark DFs and FFs, 
and presented a complete set of those model independent relations, which are
often called e.o.m. relations and the Lorentz invariance relations (LIR)\,\cite{Kanazawa:2015ajw}.  These relations allow one to express the intrinsic and kinematical
twist-3 DFs and FFs in terms of the twist-2 functions and the dynamical twist-3 
functions.  
In addition, they play a critical role to guarantee the gauge invariance and frame independence
of the twist-3 cross sections\,\cite{Kanazawa:2013uia,Kanazawa:2015ajw,Kanazawa:2015jxa}.  
In this paper, we extend the study to gluonic twist-3 DFs and FFs for a spin-1/2 
hadron\,\cite{Ji:1992eu,Mulders:2000sh,Gamberg:2018fwy,Yabe:2019awq,Kenta:2019bxd}
and derive all of those exact relations.  
For the twist-3 gluon DFs in the transversely polarized nucleon, 
which are relevant to SSAs in $ep^\uparrow\to eDX$\,\cite{Beppu:2010qn}, 
$p^\uparrow p\to DX$\,\cite{Koike:2011b}, $p^\uparrow p\to \{\gamma,\gamma^*\}X$\,\cite{Koike:2011nx}
and also $A_{LT}$ for $\vec{p} p^\uparrow\to DX$\,\cite{Hatta:2013wsa}, 
one of the present authors (S.Y.) 
derived such relations\cite{Hatta:2012jm}, 
while no such systematic studies exist for the twist-3 gluon FFs.  
There are several purely gluonic twist-3 FFs for a transversely polarized
spin-1/2 hadron, so the present study is particularly important for the study of
their contribution to the polarized 
hyperon production in the unpolarized proton-proton 
collision ($pp\to\Lambda^\uparrow X$)\,\cite{Yabe:2019awq,Kenta:2019bxd}
and SIDIS ($ep\to e\Lambda^\uparrow X$), etc.  Those exact
relations for the gluonic DFs and FFs need to be taken into account
in the derivation of the cross section and will be crucial to 
guarantee gauge invariance and the frame independence of the twist-3 cross sections
as in the case of quark DFs and FFs.   

The remainder of this paper is organized as follows:
In section 2, we derive the relations among the twist-3 gluon DFs.  After summarizing the
complete set of purely gluonic distributions up to twist-3, we derive all the constraint relations 
among those functions.  
In section 3, we extend the study to the twist-3 gluon FFs.
There are more number of twist-3 FFs compared to the twist-3 DFs
due to the lack of a constraint from time reversal invariance.  
In particular, the dynamical FFs
become complex, and the real and imaginary parts obey different
constraint relations.  
Section 4 will be devoted to a brief summary.

\section{Twist-3 gluon distributions}

\subsection{Intrinsic, kinematical and dynamical twist-3 gluon distributions}

We first summarize the definition of three types of purely gluonic distribution functions 
in the nucleon which has mass $M$, momentum $P$ ($P^2=M^2$) and the spin vector $S$ ($S^2=-M^2$).  
As usual, we introduce two light-like vectors $p$ and $n$, which satisfy
$P^\mu =p^\mu +{M^2\over 2}n^\mu$ and $p\cdot n=1$.  Below we work in  frame where 
$p^\mu=P^+g^{\mu}_{ +}$ and $n^\mu=(1/p^+) g^{\mu}_{ -}$.  
The simplest collinear gluon distribution functions are defined from the lightcone correlation functions  
of gluon's field strengths $F_a^{\alpha\nu}$
with color index $a$ 
in the nucleon
state $| PS\ra$ as\,\cite{Ji:1992eu,Mulders:2000sh}
\beq
&&\Phi^{\alpha\nu,\beta\mu}(x)=\int {d\lambda\over 2\pi}e^{i\lambda x}\la PS|F_a^{\alpha\nu}(0)[0,\lambda n]F_a^{\beta\mu}(\lambda n)|PS\ra\nonumber\\
&&\qquad={1\over 2}
\left\{ \left(-g_\perp^{\alpha\beta}p^\mu + g_\perp^{\alpha\mu}p^\beta\right)p^{\nu}
- \left( -g_\perp^{\nu\beta}p^\mu + g_\perp^{\nu\mu}p^\beta\right)p^\alpha\right\}xG(x)\nonumber\\
&&\qquad+{i\over 2}(S\cdot n)\left\{ \left( \epsilon^{pn\alpha\beta}p^\mu-\epsilon^{pn\alpha\mu}p^\beta\right)p^\nu
-\left(\epsilon^{pn\nu\beta}p^\mu-\epsilon^{pn\nu\mu}p^\beta\right)p^\alpha\right\}x\Delta G(x)\nonumber\\
&&\qquad-{iM\over 2}\left\{\left(\epsilon^{\alpha\beta nS_\perp}p^\mu - \epsilon^{\alpha\mu n S_\perp}p^\beta\right)p^\nu
-\left(\epsilon^{\nu\beta nS_\perp}p^\mu -\epsilon^{\nu\mu n S_\perp}p^\beta\right)p^\alpha\right\}x \Delta G_{3T}(x)\nonumber\\
&&\qquad +{iM\over 2}\left\{\epsilon^{\beta\mu pn}\left(S_\perp^\alpha p^\nu-S_\perp^\nu p^\alpha\right)
-\epsilon^{\alpha\nu pn}\left(S_\perp^\beta p^\mu - S_\perp^\mu p^\beta\right)\right\} x \Delta H_{3T}(x), 
\label{PDFintrinsic}
\eeq
where $g_\perp^{\alpha\beta}=g^{\alpha\beta}-p^\alpha n^\beta - p^\beta n^\alpha$, 
and the transverse spin vector $S_\perp^\mu$ is defined as $S^\mu=(S\cdot n)p^\mu+ (S\cdot p)n^\mu + MS_\perp^\mu$. 
$[0,\lambda n]\equiv P\,{\rm exp}\{ig\int_\lambda^0 dt\,A(t n)\cdot n\}$ is the gauge link which guarantees
gauge invariance of the correlation function. 
Here and below
we use the shorthand notation 
$\epsilon^{pn\alpha\beta}\equiv\epsilon^{\nu\mu\alpha\beta}p_\nu n_\mu$, {\it etc}. 
$G(x)$ and $\Delta G(x)$ are, respectively,  twist-2 unpolarized and helicity distributions and $\Delta G_{3T}(x)$ and $\Delta H_{3T}(x)$ are
the {\it intrinsic} twist-3 distributions corresponding, respectively, to $\la F^{+\perp}F^{+-}\ra$ and $\la F^{+\perp}F^{\perp\perp}\ra$
correlators.  Although $\Delta H_{3T}(x)$ drops from the correlator $\Phi^{\alpha n,\beta n}(x)$ 
which contribute to a cross section,
we need the form (\ref{PDFintrinsic}) to derive a constraint relations among the twist-3 distributions.  
Each function in (\ref{PDFintrinsic}) has a support on $|x|<1$.

The second type of the twist-3 gluon distributions are the {\it kinematical} ones which are defined as 
\beq
&&\Phi_\partial^{\alpha\beta\gamma}(x)=\int{d\lambda\over 2\pi}e^{i\lambda x}
\la PS | F_a^{\alpha n}(0) F_a^{\beta n}(\lambda n) |PS\ra i\overleftarrow{\partial}^\gamma\nonumber\\
&&
\ \equiv \lim_{\xi\to 0}
\int{d\lambda\over 2\pi}e^{i\lambda x}
\la PS | \left( F^{\alpha n}(0)[0,\infty n]\right)_a i{d\over d\xi_\gamma} \left([\infty n 
+\xi, \lambda n+\xi] F^{\beta n}(\lambda n+\xi)\right)_a |PS\ra
\nonumber\\
&&\ =
{M\over 2}g_\perp^{\alpha\beta}\epsilon^{pnS_\perp \gamma} G_T^{(1)}(x) -{iM\over 2}\epsilon^{pn\alpha\beta}S_\perp^\gamma
\Delta G_T^{(1)}(x)
+{M\over 8}\left( \epsilon^{pnS_\perp \{\alpha}g_\perp^{\beta\}\gamma}+\epsilon^{pn\gamma\{\alpha}S_\perp^{\beta\}}
\right)\Delta H_T^{(1)}(x)
\nonumber\\
&&\qquad +({\rm terms\ proportional\ to}\ p^\gamma)+\cdots,  
\label{PDFkine}
\eeq
where $\cdots$ denotes twist-4 or higher.  
These three kinematical distributions 
$\Delta G_T^{(1)}(x)$, $\Delta G_T^{(1)}(x)$ and $\Delta H_T^{(1)}(x)$
can be also written as the $k_\perp^2/M^2$-moment of the transverse momentum dependent (TMD) distributions.
Note that the TMD distribution corresponding to $\Delta G_T^{(1)}(x)$ is naively $T$-even, while those
for $G_T^{(1)}(x)$ and $\Delta H_T^{(1)}(x)$ are naively $T$-odd.  
Each function in (\ref{PDFkine}) has a support on $|x|<1$.

The third type of distributions are the dynamical ones which are defined as the lightcone 
correlation functions of 
three field strengths (``$F$-type'' distribution)\,\cite{Ji:1992eu,Beppu:2010qn}
\footnote{We follow the convention of \cite{Beppu:2010qn}.}:  
\beq
&&N_F^{\alpha\beta\gamma}(x_1,x_2)=i\int{d\lambda\over 2\pi}\int{d\mu\over 2\pi}e^{i\lambda x_1}
e^{i\mu(x_2-x_1)}\la PS| if^{acb}F_a^{n\alpha}(0)gF_c^{n\gamma}(\mu n) 
F_b^{n\beta} (\lambda n)|PS\ra\nonumber\\
&&\ 
=2iM\left[ -g_\perp^{\alpha\beta}\epsilon^{\gamma pnS_\perp}N(x_1,x_2)
+g_\perp^{\alpha\gamma}\epsilon^{\beta pnS_\perp}N(x_2,x_2-x_1)
+g_\perp^{\beta\gamma}\epsilon^{\alpha pnS_\perp}N(x_1,x_1-x_2)\right]+\cdots,\qquad
\label{PDFFtype}
\eeq
where $f^{acb}$ is the anti-symmetric structure constant for color SU(N)
and $\cdots$ denotes twist-4 or higher.  
Here and below we often suppress the gauge link operators between the field strengths for simplicity. 
$N(x_1,x_2)$ satisfies the symmetry relation $N(x_1,x_2)=N(x_2,x_1)$ and $N(-x_1,-x_2)=-N(x_1,x_2)$ and
has a support on $|x_{1,2}|<1$ and $|x_1-x_2|<1$.    
Replacing $if^{abc}$ by $d^{abc}$ (symmetric structure constants)
in $N^{\alpha\beta\gamma}_F(x_1,x_2)$, one can define another 3-gluon correlation functions.  However, 
we shall not consider them, since they are not related to any other types of  twist-3 gluon distributions.   
We call $N(x_1,x_2)$ (and $G_F(x_1,x_2)$ in (\ref{qgcorr2}) below) {\it dynamical} twist-3 DFs.
Replacing $F_c^{n \gamma}(\mu n)$ by the covariant derivative $D^\gamma(\mu n)=\partial^\gamma-igA^\gamma(\mu n)$ in (\ref{PDFFtype}), one 
obtains ``$D$-type'' distributions
as
\beq
&&N_D^{\alpha\beta\gamma}(x_1,x_2)=i\int{d\lambda\over 2\pi}\int{d\mu\over 2\pi}e^{i\lambda x_1}
e^{i\mu(x_2-x_1)}\la PS| F_a^{n\alpha}(0) \left(D^{\gamma}(\mu n) F^{n\beta} (\lambda n)\right)_a|PS\ra\nonumber\\
&&\ 
=2iM\left[ -g_\perp^{\alpha\beta}\epsilon^{\gamma pnS_\perp}D_1(x_1,x_2)
-g_\perp^{\alpha\gamma}\epsilon^{\beta pnS_\perp}D_2(x_1,x_2)
+g_\perp^{\beta\gamma}\epsilon^{\alpha pnS_\perp}D_2(x_2,x_1)\right]
\label{PDFDtype}
\\
&&\quad-{1\over 2}g_\perp^{\alpha\beta}p^\gamma G(x_1,x_2)x_2^2+{i\over 2}\epsilon^{\alpha\beta pn}p^\gamma(S\cdot n)\Delta G(x_1,x_2)x_2^2
-{i\over 2}M\epsilon^{\alpha\beta nS_\perp}p^\gamma \Delta G_{3T}(x_1,x_2)x_2^2+
\cdots,\nonumber
\eeq
where $\cdots$ denotes twist-4 or higher.  
It is easy to see that $\int dx_1N_D^{\alpha\beta n}(x_1,x)$ 
is reduced to $\Phi^{\alpha n,\beta n}(x)$, and
the three distributions in the last three terms of (\ref{PDFDtype}) are thus related to
those in (\ref{PDFintrinsic}) as
\beq
\int_{-1}^1\,dx_1f(x_1,x)=f(x),\qquad {\rm for}\ f=G,\ \Delta G,\ \Delta G_{3T}.
\eeq
Equations (\ref{PDFintrinsic}), (\ref{PDFkine}), (\ref{PDFFtype}) and (\ref{PDFDtype}) define all necessary
collinear twist-3 gluonic distribution functions in the collinear twist-3 formalism.  Below we shall derive all constraint 
relations among those functions.

\subsection{Relations between $D$- and $F$-type DFs and QCD equation of motion}
Using the identity
\beq
D^\gamma(\mu n)[\mu n,\lambda n]=ig\int_\lambda^\mu 
dt\,[\mu n,tn]F^{\gamma n}(tn)[tn,\lambda n]+[\mu n, \lambda n]D^\gamma (\lambda n),
\label{shiftD}
\eeq
$D$- and $F$- type 3-gluon correlators in (\ref{PDFDtype}) and (\ref{PDFFtype})  are connected as
\beq
N_D^{\alpha\beta\gamma}(x_1,x_2)={\cal P}{1\over x_2-x_1}N_F^{\alpha\beta\gamma}(x_1,x_2)+\delta\left(x_1-x_2\right)
\widetilde{\Phi}_\partial^{\alpha\beta\gamma}(x_1), 
\label{PDFDFrelation}
\eeq
where
\beq
&&\widetilde{\Phi}_\partial^{\alpha\beta\gamma}(x)=i\int{d\lambda\over 2\pi}e^{i\lambda x}
\la PS |F^{n\alpha}(0)[0,\lambda n]D^\gamma (\lambda n) F^{n\beta}(\lambda n)| PS\ra\nonumber\\
&&\qquad\qquad\qquad+\int {d\lambda \over 2\pi}  e^{i\lambda x} \int_{-\infty}^\infty{d\mu}   {1\over 2}\epsilon(\mu-\lambda)
\la PS | F^{n\alpha}(0)gF^{n\gamma}(\mu n) F^{n\beta}(\lambda n)|PS \ra,
\eeq
with $\epsilon(\mu-\lambda)=2\theta(\mu-\lambda)-1$.  
On the other hand, 
the correlator for the kinematical twist-3 
distributions $\Phi_\partial^{\alpha\beta\gamma}(x)$ 
in (\ref{PDFkine}) can be rewritten as
\beq
&&\Phi_\partial^{\alpha\beta\gamma}(x)=i\int{d\lambda\over 2\pi}e^{i\lambda x}
\la PS |F^{n\alpha}(0)[0,\lambda n]D^\gamma (\lambda n) F^{n\beta}(\lambda n)| PS\ra\nonumber\\
&&\qquad\qquad\qquad+\int {d\lambda \over 2\pi}  e^{i\lambda x} \int_\lambda^\infty{d\mu}  
\la PS | F^{n\alpha}(0)gF^{n\gamma}(\mu n) F^{n\beta}(\lambda n)|PS \ra.  
\eeq
Here and below we often suppress the color indices and gauge links for simplicity.  
For example
$F^{n\alpha}(0)[0,\lambda n]D^\gamma (\lambda n)F^{n\beta}(\lambda n)$
denotes $F_a^{n\alpha}(0)([0,\lambda n]D^\gamma 
(\lambda n)F^{n\beta}(\lambda n))_a$, and\\
$F^{n\alpha}(0)gF^{n\gamma}(\mu n) F^{n\beta}(\lambda n)$
represents
$if^{acb}F_a^{n\alpha}(0)gF_c^{n\gamma}(\mu n)F_b^{n\beta}(\lambda n)
\equiv\\
if^{acb}F_a^{n\alpha}(0)([0,\mu n]gF^{n\gamma}(\mu n)[\mu n,\lambda n])_c
F_b^{n\beta}(\lambda n)$.  
One thus obtain the relation 
\beq
\widetilde{\Phi}_\partial^{\alpha\beta\gamma}(x)=
\Phi_\partial^{\alpha\beta\gamma}(x)+{i\pi}N_F^{\alpha\beta\gamma} (x,x), 
\eeq
where $N_F^{\alpha\beta\gamma} (x,x)$ defines soft-gluon-pole functions.
By comparing real and imaginary parts of both sides of (\ref{PDFDFrelation}), one obtains the following relations:
\beq
&&D_1(x_1,x_2)={\cal P}{1\over x_2-x_1}N(x_1,x_2)
\label{DF1},\\
&&D_2(x_1,x_2)={\cal P}{-1\over x_2-x_1}N(x_2,x_2-x_1)-{1\over 4}\delta(x_1-x_2)\Delta G_T^{(1)}(x_1),
\label{DF2}\\
&&G_T^{(1)}(x)=4\pi\left(N(x,x)-N(x,0)\right),
\label{DF3}\\
&&\Delta H_T^{(1)}(x)=-8\pi N(x,0).  
\label{DF4}
\eeq
The first two relations were derived in \cite{Hatta:2012jm}.  
They show the 
$D$-type functions are determined by the $F$-type and kinematical functions.
The last two relations (\ref{DF3}) and (\ref{DF4}) are the analogues of the relations
for the quark distributions that show the $k_\perp^2$-moment of the ``naively $T$-odd''
TMD distribution functions, such as Sivers and Boer-Mulders functions, are
proportional to the soft-gluon-pole (SGP) function of the $F$-type quark-gluon correlation function.  It has been shown that the
SGP functions 
$N(x,x)$ and $N(x,0)$ contribute to 
SSAs for $ep^\uparrow\to e DX$\,\cite{Beppu:2010qn},
$p^\uparrow p\to DX$\,\cite{Koike:2011b}, 
$p^\uparrow p\to \{\gamma,\gamma^*\}X$\,\cite{Koike:2011nx} and
$p^\uparrow p\to \pi X$\,\cite{Beppu:2013uda}.  
To the best of our knowledge, 
the relations (\ref{DF3}) and (\ref{DF4}) were not explicitly written in the literature.  

To get further relations, we multiply $g^{\perp}_{\beta\gamma}$ to (\ref{PDFDFrelation}), integrate over $x_1$
and use the
relation
$D_{\perp\,\beta}(\lambda n)F^{n\beta}(\lambda n)=-D^n(\lambda n)F^{np}(\lambda n)+g\bar{\psi}(\lambda n)\nslash t^a\psi(\lambda n)$
which follows from the QCD equation of motion (e.o.m.), $\left( D_\mu F^{\mu\alpha}\right)_a= -g\bar{\psi}\gamma^\alpha t^a\psi$. One then obtains
\beq
&&{x^2\over 2}\Delta G_{3T}(x)+D_g(x)\nonumber\\
&&\quad=2\int_{-1}^1dx_1{\cal P}{1\over x-x_1}\left[ -N(x_1,x)+N(x,x-x_1)+2N(x_1,x_1-x)\right] -{1\over 2}\Delta G_T^{(1)}(x),
\label{EOMrelation}
\eeq
where $D_g(x)$ is defined as
\beq
M D_g(x)\epsilon^{\alpha pn S_\perp}=
\int{d\lambda\over 2\pi}\,e^{i\lambda x}\la PS | F_a^{n\alpha}(0)\bar{\psi}(\lambda n)\nslash t^a \psi(\lambda n)
|PS \ra.
\label{qgcorr1}
\eeq
$D_g(x)$ is related to the twist-3 quark-gluon correlation function 
$G_F(x_1,x_2)$ defined by\footnote{We follow the convention of 
\cite{Eguchi:2006qz,Eguchi:2006mc} for $G_F(x_1,x_2)$.}
\beq
\int{d\lambda\over 2\pi}\int{d\mu\over 2\pi}\,e^{i\lambda x_1}e^{i\mu(x_2-x_1)}\la PS | \bar{\psi}(0) F_a^{\alpha n}(\mu n)\nslash t^a \psi(\lambda n)
|PS \ra =M\epsilon^{\alpha pn S_\perp} G_F(x_1,x_2),
\label{qgcorr2}
\eeq
as
\beq
D_g(x)=-\int_{-1}^1dx_1G_F(x_1,x_1-x),
\eeq
with the support on $|x|<1$. 
The relation (\ref{EOMrelation}) is also new.  

From (\ref{PDFDFrelation}), one can obtain another relation involving $\Delta H_{3T}(x)$ as follows.
We first write
\beq
\Phi^{\alpha n,\beta\mu}(x)&=&{-i\over x}
\int{d\lambda \over 2\pi}{d e^{i\lambda x}\over d\lambda}
\la PS | F^{\alpha n}(0)[0,\lambda n] F^{\beta\mu}(\lambda n)| PS\ra\nonumber\\
&=&{i\over x}\int{d\lambda \over 2\pi}e^{i\lambda x}\la PS | F^{\alpha n}(0)[0,\lambda n]  D^n(\lambda n) F^{\beta\mu}(\lambda n)| PS\ra,
\label{PDFDF2}
\eeq
where we have used the relation ${d\over d\lambda}[0,\lambda n]F^{\beta\mu}(\lambda n)=
[0,\lambda n]D^n(\lambda n)F^{\beta\mu}(\lambda n)$ after integration by part.  
We then use the Bianchi identity $D^nF^{\beta\mu}=-D^\beta F^{\mu n}+D^\mu F^{\beta n}$ to get
\beq
\Phi^{\alpha n,\beta\mu}(x)={1\over x}\int dx_1\left\{ N_D^{\alpha\beta\mu}(x_1,x)- N_D^{\alpha\mu\beta }(x_1,x)\right\}.  
\label{DFrelation3}
\eeq
Taking $\alpha$, $\beta$ and $\mu$ to be transverse, one arrives at the following relation
\beq
x\Delta H_{3T}(x)={4\over x}\int dx_1\left\{ D_1(x_1,x)-D_2(x_1,x)\right\}.
\eeq
Using the relations, (\ref{DF1}) and (\ref{DF2}), in this equation
one eventually obtains
\beq
{1\over 2}x^2\Delta H_{3T}(x)=2\int dx_1{\cal P}{1\over x-x_1}\left\{ N(x_1,x)+N(x,x-x_1)\right\} + {1\over 2}\Delta G_T^{(1)}(x).
\label{DFrelation2}
\eeq
This relation was derived here for the first time.  

To summarize this subsection, we have obtained two relations (\ref{EOMrelation}) and (\ref{DFrelation2})
which relate the two intrinsic functions, $\Delta G_{3T}(x)$ and $\Delta H_{3T}(x)$, and one kinematical
function, $\Delta G_T^{(1)}(x)$, to the dynamical functions.  
One needs another independent
relation to express those three functions in terms of the dynamical functions.  

\subsection{Constraint relations from nonlocal operator product expansion}
Here we derive a relation from the nonlocal version of the
operator product expansion (OPE) for
general correlation functions not necessarily on the lightcone.  The method was originally developed in 
\cite{Balitsky:1987bk,Balitsky:1989ry}, and have been frequently 
used for the twist-3 distributions\,\cite{Balitsky:1987bk,
Kodaira:1998jn,Eguchi:2006qz,Hatta:2012jm,Braun:2001qx},
the twist-3 fragmentation functions\,\cite{Balitsky:1990ck,Kanazawa:2015ajw}, 
and the distribution amplitudes for hard exclusive
processes\,\cite{Balitsky:1989ry,Braun:1989iv,Ball:1998sk}, etc.
This method is 
equivalent to OPE, and incorporates all the constraints
from Lorentz invariance property of the correlation functions. 
Here we apply this method to the twist-3 gluon distribution functions to
derive constraint relations.  

We start from the following operator identity: 
\beq
&&
{\partial \over \partial y_\rho}\left[ F^{\alpha\nu}(-y)[-y,y]F^{\beta\mu}(y)\right]
\nonumber\\
&&
\qquad=-F^{\alpha\nu}(-y)\overleftarrow{D}^\rho(-y)[-y,y]F^{\beta\mu}(y)
+F^{\alpha\nu}(-y)[-y,y]\overrightarrow{D}^\rho(y)F^{\beta\mu}(y)
\nonumber\\
&&
\qquad+i\int_{1}^{-1}dt\,t\,F^{\alpha\nu}(-y)[-y,ty]gF^{\rho y}(ty)[ty,y]F^{\beta\mu}(y).
\label{identity1}
\eeq
In the left hand side (l.h.s.) of this equation, one should first make $y$ non-lightlike, and take the
lightcone limit $y^\mu\to y^- g^{\mu}_-$ after taking the derivative.  
From translational invariance, we have another identity,
\beq
&&0=\lim_{\xi\to0}{d\over d\xi_\rho} \la PS|  F^{\alpha\nu}(-y+\xi)[-y+\xi,y+\xi]F^{\beta\mu}(y+\xi)|PS\ra\nonumber\\
&&=\la PS|F^{\alpha\nu}(-y)\overleftarrow{D}^\rho(-y)[-y,y]F^{\beta\mu}(y)|PS\ra
+\la PS|F^{\alpha\nu}(-y)[-y,y]\overrightarrow{D}^\rho(y)F^{\beta\mu}(y)|PS\ra\nonumber\\
&&\qquad+i\int_{1}^{-1}dt\,\la PS|F^{\alpha\nu}(-y)[-y,ty]gF^{\rho y}(ty)[ty,y]F^{\beta\mu}(y)|PS\ra.
\label{identity2}
\eeq 
We take the expectation value of (\ref{identity1}) by $|PS\ra$, and use 
(\ref{identity2}) to eliminate the first term in the right hand side (r.h.s.).  We then obtain
\beq
&&{\partial \over \partial y_\rho}\la PS|
F^{\alpha\nu}(-y)[-y,y]F^{\beta\mu}(y)|PS \ra\nonumber\\
&&=2
\la PS| F^{\alpha\nu}(-y)[-y,y]\overrightarrow{D}^\rho(y)F^{\beta\mu}(y)\|PS \ra\nonumber\\
&&\qquad+\la PS| i\int_{1}^{-1}dt\,(t+1)\,F^{\alpha\nu}(-y)[-y,ty]gF^{\rho y}(ty)[ty,y]F^{\beta\mu}(y)|PS\ra.  
\eeq
From this equation, one obtains the identity
\beq
&&
{\partial \over \partial y^\beta}\la PS | F^{\alpha y}(-y)[-y,y] F^{\beta y}(y)| PS \ra\nonumber\\
&&\qquad=\la PS | F^{\alpha}_{\ \,\mu}(-y)[-y,y]F^{\mu y}(y)|PS\ra+2\la PS| F^{y\alpha}(-y)
g\bar{\psi}(y)\yslash t^a \psi(y)|PS\ra\nonumber\\
&&\qquad\qquad+\la PS| i\int_{1}^{-1}dt\,(t+1)\,F^{\alpha y}(-y)[-y,ty]gF_{\beta}^{\ y}(ty)[ty,y]
F^{\beta y}(y)|PS\ra, 
\label{correlator1}
\eeq
where we used the QCD e.o.m., $D_\beta(y)F^{\beta\mu}(y)=-g\bar{\psi}(y)\gamma^\mu t^a \psi(y)$, 
in the second term of r.h.s.
In order to get a relation among the twist-3 distributions from (\ref{correlator1}), one needs 
inverse Fourier transform of (\ref{PDFintrinsic}), (\ref{PDFFtype}) and (\ref{qgcorr2}).  
In particular, to calculate l.h.s. and the first term in the r.h.s. of (\ref{correlator1}), 
one has to use the following form: 
\beq
&&\la PS | F^{\alpha\nu}(-y)[-y,y]F^{\beta\mu}(y) | PS\ra\nonumber\\
&&\quad=\int dx\,e^{-2ixp\cdot y}\left[
{1\over 2}\left\{\left( -g_\perp^{\alpha\beta}p^\mu+g_\perp^{\alpha\mu}p^\beta\right) p^\nu
-\left( -g_\perp^{\nu\beta}p^\mu+g_\perp^{\nu\mu}p^\beta\right) p^\alpha\right\} xG(x)\right.\nonumber\\
&&\qquad+\left. {iS\cdot y\over 2(p\cdot y)^2}
\left\{
\left(\epsilon^{py\alpha\beta}p^\mu- \epsilon^{py\alpha\mu}p^\beta\right)p^\nu - 
\left(\epsilon^{py\nu\beta}p^\mu- \epsilon^{py\nu\mu}p^\beta\right)p^\alpha
\right\} x\Delta G(x)
\right.\nonumber\\
&&\qquad\left.-{iM\over 2p\cdot y }\left\{ 
\left( \epsilon^{\alpha\beta y S_\perp}p^\mu -\epsilon^{\alpha\mu y S_\perp} p^\beta\right)p^\nu
-\left( \epsilon^{\nu\beta y S_\perp}p^\mu -\epsilon^{\nu\mu y S_\perp} p^\beta\right)p^\alpha\right\} x \Delta G_{3T}(x)\right.\nonumber\\
&&\qquad\left.+{iM\over 2p\cdot y}
\left\{
\epsilon^{\beta\mu p y}\left( S_\perp^\alpha p^\nu - S_\perp^\nu p^\alpha\right)
-\epsilon^{\alpha\mu p y}\left( S_\perp^\beta p^\nu - S_\perp^\nu p^\beta\right)
\right\} x\Delta H_{3T}(x)\right].
\label{PDFintrinsic-inv}
\eeq
In taking the derivative of (\ref{PDFintrinsic-inv}) with respect to $y^\beta$, one should
use the form $S_\perp^\mu = S^\mu -{S\cdot y\over p\cdot y}p^\mu$ and $g_\perp^{\mu\nu}
=g^{\mu\nu}-{p^\mu y^\nu+p^\nu y^\mu\over p\cdot y}$, keep all components of 
$y^\mu$ with $y^2\neq 0$ and then take $y^\mu\to g^{\mu}_-y^-$ limit.  
With this procedure, we have eventually obtained the following relation:
\beq
&&x{\partial \over \partial x}\left( x \Delta G_{3T}(x)\right)+ x\Delta H_{3T}(x) +x\Delta G(x)\nonumber\\
&&\qquad=-2{d D_g(x)\over dx} +4\int\,dx_1{\cal P}{1\over x-x_1}{\partial \over \partial x}\left\{ 2N(x_1,x_1-x)+N(x,x-x_1)-N(x_1,x)\right\}\nonumber\\
&&\qquad+4\int dx_1{\cal P}{1\over (x-x_1)^2}\left\{ -N(x_1,x_1-x) -2N(x,x-x_1)+N(x_1,x)\right\}.
\label{OPE1}
\eeq
This relation is independent from (\ref{EOMrelation}) and (\ref{DFrelation2}), 
and the three relations 
(\ref{EOMrelation}), (\ref{DFrelation2}) and (\ref{OPE1}) 
allow one
to solve $\Delta G_{3T}(x)$, $\Delta H_{3T}(x)$ and $\Delta G_T^{(1)}(x)$ in terms of $\Delta G(x)$ and the
dynamical functions. 

Here we comment on the relations obtained from operator identities other than (\ref{identity1}).  
One can derive a constraint relation
by considering the following correlation function:  
\beq
y_\rho\left[ {\partial \over \partial y_\rho}\la PS| F^{y\nu}(-y)[-y,y]F^{\beta}_{\ \,\nu}(y)|PS \ra - ( \rho \leftrightarrow \beta)\right].
\label{OPE2}
\eeq
We found that this correlator simply gives the relation that is obtained from 
(\ref{EOMrelation}) and (\ref{DFrelation2}), which supplies a good consistency check.  
We also found that the operator identity for the correlator
\beq
y_\rho\left[ {\partial \over \partial y_\rho}\la PS| F^{y\nu}(-y)[-y,y]\widetilde{F}^{\beta}_{\ \,\nu}(y)|PS \ra
 - ( \rho \leftrightarrow \beta)\right],
\label{hatta1}
\eeq
with $\widetilde{F}^{\beta\nu}={1\over 2}\epsilon^{\beta\nu\rho\tau}F_{\rho\tau}$ 
gives the same relation as (\ref{OPE1}), which also serves to confirm our result. 

It is interesting to compare our approach and that in \cite{Hatta:2012jm}.  
The authors of
\cite{Hatta:2012jm} analyzed the correlator (\ref{hatta1}) 
to express
$\DGhat_{3T}(x)$ and $\DGhat_T^{(1)}(x)$ in terms of $\Delta G(x)$ and the dynamical twist-3 distributions.
They started from the identity
\beq
&&y_\rho\left[ {\partial \over \partial y_\rho}\la PS| F^{y\nu}(-y)\widetilde{F}^{\beta}_{\ \,\nu}(y)|PS \ra
 - ( \rho \leftrightarrow \beta)\right]\nonumber\\
&&\qquad =\la PS|\left( F^{y\nu}(-y)\widetilde{F}^{\beta}_{\ \nu}(y)- F^{\beta\nu}(-y)\widetilde{F}^y_{\ \nu}(y)\right)|PS\ra\nonumber\\
&&\qquad\quad+y_\alpha y_\rho\left[ {\partial \over \partial y_\rho}\la PS| F^{\alpha\nu}(-y)\widetilde{F}^{\beta}_{\ \,\nu}(y)|PS \ra
 - ( \rho \leftrightarrow \beta)\right].  
\label{hatta2}
\eeq
The second term in the r.h.s. can be rewritten further to be expressed in 
terms of the $F$-type functions.
In our approach, the l.h.s. and the first term in the r.h.s. are calculated by using
(\ref{PDFintrinsic-inv}) and are expressed in terms of the intrinsic distributions.
In this method, $\Delta H_{3T}(x)$ does not 
survive in the l.h.s., while it does appear in the first term of the
r.h.s.  This procedure leads to the same relation as (\ref{OPE1}).  
As for the method of \cite{Hatta:2012jm}, they treated the l.h.s. of (\ref{hatta2}) 
in the same way as ours (although they did not refer to the
presence of  $\Delta H_{3T}$ term).
On the other hand, they analyzed the first term 
in the r.h.s. of (\ref{hatta2}) in a different way.  They did not 
use the form (\ref{PDFintrinsic-inv}), but rewrote it directly in terms of the $F$-type functions. 
Therefore they could obtain the constraint relation among
the twist-3 distribution functions without recourse to $\Delta H_{3T}(x)$ contribution at any stage.  
As we will see in the next subsection, our results for $\Delta G_{3T}(x)$ and $\Delta G_T^{(1)}(x)$
agree with those in \cite{Hatta:2012jm}.  Our approach can also supply the
expression for $\Delta H_{3T}(x)$.  (See next subsection.)

\subsection{Solution for intrinsic and kinematical DFs in terms of twist-2 and dynamical twist-3 DFs}
As we found in previous subsections, eqs. (\ref{EOMrelation}), (\ref{DFrelation2}) 
and (\ref{OPE1}) constitute a complete set of the
independent relations among the twist-3 intrinsic, kinematical and dynamical DFs.  Here
we provide a solution for the intrinsic and kinematical functions in terms of the twist-2 and 
dynamical twist-3 DFs.  
Taking the sum of (\ref{EOMrelation}) and (\ref{DFrelation2}), we obtain
\beq
x\Delta H_{3T}(x)=-x\Delta G_{3T}(x)-{2\over x}D_g(x) +{8\over x}\int dx_1{\cal P}{1\over x-x_1}
\left\{ N(x,x-x_1)+N(x_1,x_1-x)\right\}.
\label{H3G3}
\eeq
Inserting this into (\ref{OPE1}) to eliminate $\Delta H_{3T}(x)$, we have
\beq
&&x^2{d\over dx}\Delta G_{3T}(x) +x\Delta G(x)+ 2x{d\over dx}\left( {D_g(x)\over x}\right) \nonumber\\
&&\qquad =4\int dx_1 {\cal P}{1\over x-x_1}{\partial \over \partial x}
\left\{2N(x_1,x_1-x) +N(x,x-x_1)-N(x_1,x)\right\} \nonumber\\
&&\qquad+
4\int dx_1{\cal P}{1\over (x-x_1)^2}\left\{ -N(x_1,x_1-x)-2N(x,x-x_1)+N(x_1,x)\right\}\nonumber\\
&&\qquad-{8\over x}\int dx_1 {\cal P}{1\over x-x_1}\left\{ N(x, x-x_1)+N(x_1,x_1-x)\right\}.
\eeq
This equation can be integrated to give 
\beq
\Delta G_{3T}(x)&=&-\int_{\epsilon(x)}^x dx_1{\Delta G(x_1)\over x_1}-2\left\{ {D_g(x)\over x^2} 
+ \int_{\epsilon(x)}^x dx_1 {D_g(x_1)\over x_1^3}\right\}\nonumber\\
&&+{4\over x^2}\int_{-1}^1dx_1{\cal P}{1\over x-x_1}\left\{2N(x_1,x_1-x)+N(x,x-x_1)-N(x_1,x)\right\}\nonumber\\
&&+\int_{\epsilon(x)}^xdx_2{8\over x_2^3}\int_{-1}^1dx_1{\cal P}{1\over x_2-x_1}\left\{ N(x_1,x_1-x_2)-N(x_1,x_2)\right\}\nonumber\\
&&+\int_{\epsilon(x)}^x dx_2 {4\over x_2^2}\int_{-1}^1{\cal P}{1\over (x_2-x_1)^2}\left\{ N(x_1,x_1-x_2)-N(x_2,x_2-x_1)\right\}.
\label{DelG3T}
\eeq
Combining this result and (\ref{EOMrelation}), one obtains the expression for $\Delta G_T^{(1)}(x)$ as
\beq
\Delta G_T^{(1)}(x)&=&x^2\int_{\epsilon(x)}^x dx_1{\Delta G(x_1)\over x_1} + 2x^2 \int_{\epsilon(x)}^x dx_1{D_g(x_1)\over x_1^3}\nonumber\\
&&-x^2\int_{\epsilon(x)}^x dx_2 {8\over x_2^3}\int_{-1}^1dx_1{\cal P}{1\over x_2-x_1}\left\{ N(x_1,x_1-x_2)-N(x_1,x_2)\right\}\nonumber\\
&&-x^2\int_{\epsilon(x)}^x dx_2 {4\over x_2^2}\int_{-1}^1dx_1{\cal P}{1\over (x_2-x_1)^2}\left\{ N(x_1,x_1-x_2)-N(x_2,x_2-x_1)\right\}.
\label{GT1}
\eeq
The result in (\ref{DelG3T}) and (\ref{GT1}) agrees with that in \cite{Hatta:2012jm}.  
 Insertion of (\ref{DelG3T}) into (\ref{H3G3}) gives the expression for $\Delta H_{3T}(x)$ as
 \beq
 \Delta H_{3T}(x)&=&\int_{\epsilon(x)}^x dx_1 {\Delta G(x_1)\over x_1} + 2 \int_{\epsilon(x)}^x dx_1{D_g(x_1)\over x_1^3}
 \nonumber\\
 &&
 +{4\over x^2}\int_{-1}^1dx_1\,{\cal P}{1\over x-x_1}\left\{ N(x,x-x_1)+N(x_1,x)\right\}\nonumber\\
 &&-8\int_{\epsilon(x)}^x dx_2 {1\over x_2^3}\int_{-1}^1 dx_1{\cal P}{1\over x_2-x_1}\left\{ N(x_1,x_1-x_2)-N(x_1,x_2)\right\}\nonumber\\
 &&-4\int_{\epsilon(x)}^x dx_2 {1\over x_2^2}\int_{-1}^1 dx_1{\cal P}{1\over (x_2-x_1)^2}\left\{ N(x_1,x_1-x_2)-N(x_2,x_2-x_1)\right\}.  
 \label{DelH3T}
 \eeq
This result is new.  
As shown in (\ref{DelG3T}), (\ref{GT1}) and (\ref{DelH3T}), the intrinsic and kinematical twist-3  gluonic distributions
are completely determined by 
$\Delta G(x)$ (often called Wandzura-Wilczek contribution) and the
$F$-type purely gluonic correlation function $N(x_1,x_2)$ and the quark-gluon correlation function
$G_F(x_1,x_2)$.
Since these relations are model independent exact relations, 
they need to be satisfied in phenomenological applications. 
These relations also provide a basis for the renormalization of  the intrinsic and the kinematical twist-3 distributions.  
The evolution equation for $N(x_1,x_2)$ and $G_F(x_1,x_2)$ have already been 
derived in \cite{Braun:2009mi}.  
The above relations (\ref{DelG3T}), (\ref{GT1}) and (\ref{DelH3T}) shows it also determines
the scale dependence of  $\Delta G_{3T}(x)$, $\Delta G_T^{(1)}$ and $\Delta H_{3T}(x)$.

\section{Twist-3 gluon fragmentation functions}

\subsection{Intrinsic, kinematical and dynamical twist-3 gluon fragmentation functions}
In this section we extend our analysis in the previous section to the
twist-3 gluon fragmentation function (FFs).  
We consider FFs for a spin-1/2 baryon 
with mass $M_h$, four momentum $P_h$ ($P_h^2=M_h^2$), and the spin vector
$S$ ($S^2=-M_h^2$).  In the twist-3 accuracy, we can treat $P_h$ as lightlike and introduce another lightlike vector
$w$ by the relation $P_h\cdot w=1$.  
We again work in a frame where $P_h^\mu = P_h^+g^{\mu}_{+}$ and 
$w^\mu=g^\mu_{-}/P_h^+$.   Transverse spin vector for the
baryon $S_\perp^\mu$ is normalized as $S_\perp^2=-1$.  
Similarly to (\ref{PDFintrinsic}), the gluon's collinear FFs
can be defined from the following fragmentation matrix elements\cite{Mulders:2000sh}:
\beq
&&\hspace{-0.3cm} \widehat{\Gamma}^{\alpha\nu,\beta\mu}(z)=
{1\over N^2-1}\int {d\lambda\over 2\pi} e^{-i\lambda/z}\sum_X \la 0| 
\left([\infty w, 0]F^{\alpha\nu}(0)\right)_a |hX\ra\la hX|
\left(F^{\beta\mu}(\lambda w)[\lambda w, \infty w]\right)_a |0\ra\nonumber\\
&&=
\left\{ 
\left( -g_\perp^{\alpha\beta}P_h^\mu+g_\perp^{\alpha\mu}P_h^\beta \right)P_h^\nu -
\left( -g_\perp^{\nu\beta}P_h^\mu+g_\perp^{\nu\mu}P_h^\beta \right)P_h^\alpha 
\right\}
\Ghat(z)
\nonumber\\
&&+i(S\cdot w)
\left\{ \left( 
\epsilon^{P_h w \alpha \beta}P_h^\mu -  \epsilon^{P_h w \alpha \mu}P_h^\beta \right)P_h^\nu 
-
\left( 
\epsilon^{P_h w \nu \beta}P_h^\mu -  \epsilon^{P_h w \nu \mu}P_h^\beta \right)P_h^\alpha 
\right\} \DGhat(z)\nonumber\\
&&
-M_h \left[
\left\{ 
\left(w^\alpha\epsilon^{\beta P_h w S_\perp } + w^\beta\epsilon^{\alpha P_h w S_\perp } \right) P_h^\mu 
-
\left(w^\alpha\epsilon^{\mu P_h w S_\perp } + w^\mu \epsilon^{\alpha P_h w S_\perp } \right) P_h^\beta
\right\} P_h^\nu\right.\nonumber\\
&&
\left. \quad\ \ -
\left\{ 
\left(w^\nu\epsilon^{\beta P_h w S_\perp } + w^\beta\epsilon^{\nu P_h w S_\perp } \right) P_h^\mu 
-
\left(w^\nu\epsilon^{\mu P_h w S_\perp } + w^\mu \epsilon^{\nu P_h w S_\perp } \right) P_h^\beta
\right\} P_h^\alpha\
\right]\DGhat_{\3Tb} (z)\nonumber\\
&&-iM_h
\left\{ 
\left( \epsilon^{\alpha \beta w S_\perp}P_h^\mu -\epsilon^{\alpha\mu w S_\perp} P_h^\beta\right) P_h^\nu
-
\left( \epsilon^{\nu \beta w S_\perp}P_h^\mu -\epsilon^{\nu \mu w S_\perp} P_h^\beta\right) P_h^\alpha
\right\} \DGhat_{3T}(z)
\nonumber\\
&&
+M_h \left\{ 
\epsilon^{\beta\mu P_h w}\Bigl( S_\perp^\alpha P_h^\nu - S_\perp^\nu P_h^\alpha \Bigr)
+
\epsilon^{\alpha\nu P_h w}\left( S_\perp^\beta P_h^\mu - S_\perp^\mu P_h^\beta \right)
\right\} \DHhat_{\3Tb}(z)
\nonumber\\
&&
+i M_h \left\{ 
\epsilon^{\beta\mu P_h w}\Bigl( S_\perp^\alpha 
P_h^\nu - S_\perp^\nu P_h^\alpha \Bigr)
-
\epsilon^{\alpha\nu P_h w}\left( S_\perp^\beta P_h^\mu - S_\perp^\mu P_h^\beta \right)
\right\} \DHhat_{3T}(z)+\cdots, 
\label{FF1}
\eeq
where $N=3$ is the number colors for $SU(N)$ and 
$+\cdots$ denotes twist-4 or higher.  All functions in (\ref{FF1}) are defined as real.  
Note that
the last two terms drop in the correlator $\sim \la F^{w\nu}\ra\la F^{w\mu}\ra$, but we need
this general correlator to derive relations among the twist-3 gluonic FFs. 
$\widehat{G}(z)$ and $\Delta\widehat{G}(z)$ are, respectively, twist-2 unpolarized and helicity FFs, and 
other 4 functions $\Delta\Ghat_{3T}(z)$, $\Delta\Ghat_{\3Tb}(z)$,  
$\Delta\Hhat_{3T}(z)$ and  $\Delta\Hhat_{\3Tb}(z)$ are intrinsic twist-3 FFs.
Compared with the distribution functions, the number of twist-3 FFs is doubled
due to the absence of the constraint from time reversal invariance, i.e., ``naively $T$-odd'' FFs
$\Delta\Ghat_{\3Tb}(z)$ and $\Delta\Hhat_{\3Tb}(z)$ survive
in addition to ``naively $T$-even''
$\Delta\Ghat_{3T}(z)$ and $\Delta\Hhat_{3T}(z)$.  
Each function in (\ref{FF1}) has a support on $0<z<1$.  

The second type of gluon's FFs are the kinematical FFs, which are 
defined by 
\beq
&&\widehat{\Gamma}_\partial^{\alpha\beta\gamma} (z)={1\over N^2-1}
\sum_X\int{d\lambda\over 2\pi}e^{-i\lambda/z}\la 0|\left( [\infty w, 0]F^{w\alpha}(0)\right)_a|hX\ra 
\la hX|\left(F^{w\beta}(\lambda w)[\lambda w,\infty w]\right)_a|0\ra
i\overleftarrow{\partial}^\gamma\nonumber\\
&&\qquad=-{M_h\over 2}g_\perp^{\alpha\beta}\epsilon^{\gamma P_hwS_\perp}\Ghat_T^{(1)}(z)
-i{M_h\over 2}\epsilon^{\alpha\beta P_h w}S_\perp^\gamma\DGhat_T^{(1)}(z)\nonumber\\
&&\qquad+{M_h\over 8}\left( \epsilon^{P_h wS_\perp\{ \alpha}g_\perp^{\beta\}\gamma} +
\epsilon^{P_h w\gamma\{\alpha}S_\perp^{\beta\}}
\right)\DHhat_T^{(1)}(z)
+({\rm terms\ proportional\ to}\ P_h^\gamma)+\cdots,
\label{FFkine}
\eeq
where $\cdots$ denotes twist-4 or higher. 
These three kinematical FFs $\Ghat_T^{(1)}(z)$, $\DGhat_T^{(1)}(z)$ and $\DHhat_T^{(1)}(z)$ can also 
be written as the $k_\perp^2/M_h^2$-moment of the TMD FFs
as in (\ref{PDFkine}) for the distribution functions.  Each function has a support on $0<z<1$.  

The third type of twist-3 FFs are the dynamical ones which are defined as the three gluon correlation function\cite{Gamberg:2018fwy,Yabe:2019awq,Kenta:2019bxd}:
\beq
\widehat{N}_F^{\alpha\beta\gamma}\left({1\over z_1},{1\over z_2}\right)&&={i\over N^2-1}\sum_X
\int{d\lambda\over 2\pi}\int{d\mu\over 2\pi}e^{-i{\lambda\over z_1}}e^{-i\mu({1\over z_2}-{1\over z_1})}
if^{abc}\la 0| F_a^{w\alpha}(0)|hX\ra\nonumber\\[-5pt]
&&\qquad\qquad\qquad\qquad\qquad\times\la hX|F_b^{w\beta}(\lambda w)gF_c^{w\gamma}(\mu w)|0\ra
\nonumber\\[15pt]
&&=iM_h\left[ -g_\perp^{\alpha\beta}\epsilon^{\gamma P_h w S_\perp}\Nhat_2\left({1\over z_2}-{1\over z_1},{1\over z_2}\right)
+g_\perp^{\alpha\gamma}\epsilon^{\beta P_h w S_\perp}\Nhat_2\left({1\over z_1},{1\over z_2}\right)\right.\nonumber\\
&&\left.
\qquad\qquad\qquad\qquad+g_\perp^{\beta\gamma}\epsilon^{\alpha P_h w S_\perp}\Nhat_1\left({1\over z_1},{1\over z_2}\right)
\right], 
\label{FFFtype}
\eeq
where the color indices of the field strength are contracted by the anti-symmetric 
structure constant $i f^{abc}$ and
the presence of appropriate gauge links similar to (\ref{FF1}) is implied 
to guarantee gauge invariance of the FFs.  
There are two independent $F$-type FFs $\Nhat_1\left({1\over z_1},{1\over z_2}\right)$
and $\Nhat_2\left({1\over z_1},{1\over z_2}\right)$ which are in general {\it complex}, meaning 
that the number of independent 
$F$-type FFs is four times more than the distribution case.  
$\Re\Nhat_{1,2}\left({1\over z_1},{1\over z_2}\right)$ are naively $T$-even, while
$\Im\Nhat_{1,2}\left({1\over z_1},{1\over z_2}\right)$ are naively $T$-odd.  
$\Nhat_{1,2}\left({1\over z_1},{1\over z_2}\right)$ have a support on 
${1\over z_2}>1$ and ${1\over z_2}>{1\over z_1}>0$.  
Replacing $if^{abc}$ by 
the symmetric structure constants $d^{abc}$, one can define another $F$-type FFs.    
Although they appear in a certain cross section, 
e.g., $pp\to\Lambda^\uparrow X$\,\cite{Yabe:2019awq,Kenta:2019bxd}, 
they are not related to other twist-3 FFs.  
We therefore do not consider those FFs hereafter.

One can also define another set of
twist-3 FFs by the replacement of $gF^{w\gamma}(\mu w)\to
D^\gamma(\mu w)$ in (\ref{FFFtype}), which gives
\beq
&&
\widehat{N}_D^{\alpha\beta\gamma}\left({1\over z_1},{1\over z_2}\right)={i\over N^2-1}\sum_X
\int{d\lambda\over 2\pi}\int{d\mu\over 2\pi}e^{-i{\lambda\over z_1}}e^{-i\mu({1\over z_2}-{1\over z_1})}\nonumber\\
&&\qquad\qquad\qquad\qquad\qquad\qquad\qquad
\times\la 0| F_a^{w\alpha}(0)|hX\ra
\la hX|\left(F^{w\beta}(\lambda w)\overleftarrow{D}^{\gamma}(\mu w)\right)_a|0\ra\nonumber\\[5pt]
&&\qquad=iM_h\left[ g_\perp^{\alpha\beta}\epsilon^{\gamma P_h w S_\perp}\Dhat_2\left({1\over z_1},{1\over z_2}\right)
+g_\perp^{\alpha\gamma}\epsilon^{\beta P_h w S_\perp}\Dhat_3\left({1\over z_1},{1\over z_2}\right)\right.\nonumber\\[-5pt]
&&\qquad\left.
\qquad\qquad\qquad\qquad+g_\perp^{\beta\gamma}\epsilon^{\alpha P_h w S_\perp}\Dhat_1\left({1\over z_1},{1\over z_2}\right)
\right]\nonumber\\
&&\qquad+P_h^\gamma g_\perp^{\alpha\beta}\Ghat\left({1\over z_1},{1\over z_2}\right)
{1\over z_2}
-iP_h^\gamma\epsilon^{P_h w\alpha\beta}(S\cdot w)
\DGhat\left({1\over z_1},{1\over z_2}\right){1\over z_2}\nonumber\\
&&\qquad+iM_h P_h^\gamma \epsilon^{\alpha\beta w S_\perp}\DGhat_{3T}\left({1\over z_1},{1\over z_2}\right){1\over z_2}
-M_hP_h^\gamma \epsilon^{P_h w S_\perp\{\alpha}w^{\beta\}}
\DGhat_{\3Tb}\left({1\over z_1},{1\over z_2}\right){1\over z_2}, 
\label{FFDtype}
\eeq
where gauge links are suppressed for simplicity.  
$\Dhat_{1,2,3}\left({1\over z_1},{1\over z_2}\right)$ are also complex functions, and 
are called $D$-type FFs.  
Functions in the last two lines are related to those in (\ref{FF1}):  
From the relation
\beq
-z \int d\left({1\over z_1}\right)
\Nhat_D^{\alpha\beta\gamma}\left( {1\over z_1}, {1\over z}\right)w_\gamma
=\widehat{\Gamma}^{\alpha w,\beta w}(z),
\eeq
it is easy to see
\beq
\int d\left({1\over z_1}\right)\widehat{f}\left( {1\over z_1},{1\over z}\right)
=\widehat{f}(z),\qquad {\rm for}\ \widehat{f}
=\Ghat,\ \DGhat,\ \DGhat_{3T},\ \DGhat_{\3Tb}.
\eeq

Finally we introduce another dynamical FFs defined by
\beq
\widetilde{\Delta}^\alpha\left({1\over z_1},{1\over z_2}\right)
&=&{1\over N}\sum_X\int{d\lambda\over 2\pi}\int{d\mu\over 2\pi}e^{-i{\lambda\over z_1}}
e^{-i\mu({1\over z_2}-{1\over z_1})}
\la 0| F_a^{w\alpha}(\mu w)|hX\ra\la hX|\bar{\psi}_j(\lambda w)t^a\psi_i(0)| 
0\ra\nonumber\\
&=&M_h\left[ e^{\alpha P_hwS_\perp}(\Pslash_h)_{ij} \widetilde{D}_{FT}
\left({1\over z_1},{1\over z_2}\right)+
iS_\perp^\alpha \left(\gamma_5\Pslash_h\right)_{ij}\widetilde{G}_{FT}
\left({1\over z_1},{1\over z_2}\right)
\right],
\label{FFtilde}
\eeq
where the spinor indices $i,\,j$ are shown explicitly.  These two
functions $\widetilde{D}_{FT}$ and 
$\widetilde{G}_{FT}$ are, in general, complex functions with their naively ``$T$-even''
real part and the ``$T$-odd'' imaginary part.  
They have a support on ${1\over z_1}>0$, ${1\over z_2}<0$ and ${1\over z_1}-{1\over z_2}>1$.  
As we will see below, constraint relations for the twist-3 gluonic FFs
involve these $F$-type quark-gluon correlation functions through QCD e.o.m. 
We collectively call the functions in (\ref{FFFtype}) and (\ref{FFtilde})
{\it dynamical}
twist-3 FFs.

\subsection{Relations between $D$- and $F$-type FFs and QCD equation of motion}
The gluon FFs introduced in (\ref{FF1})-(\ref{FFDtype}) are not independent, but are
related by various operator identities.  
Using the identity (\ref{shiftD}), we find $D$-type and $F$-type FFs are related
as
\beq
\Nhat_D^{\alpha\beta\gamma}\left({1\over z_1},{1\over z_2}\right)={\cal P}\left({-1\over {1\over z_2}-{1\over z_1}}\right)
\Nhat_F^{\alpha\beta\gamma}\left({1\over z_1},{1\over z_2}\right)+\delta\left(  {1\over z_2}-{1\over z_1}\right)
\widehat{\Gamma}_\partial^{\alpha\beta\gamma}(z_1).  
\label{FFDFrelation}
\eeq
An important difference of this relation from the similar one for the distribution function
(\ref{PDFDFrelation}) is that the correlator for the kinematical FFs appear
directly as the coefficient of $\delta$-function.
This is because $F$-type FFs become 0 at $z_1=z_2$ due to the support property
as shown in \cite{Meissner:2008yf,Gamberg:2008yt}. 
From (\ref{FFDFrelation}), we have
\beq
&&\Dhat_1\left({1\over z_1},{1\over z_2}\right)={\cal P}{1\over {1\over z_2}-{1\over z_1}} \Nhat_1\left({1\over z_1},{1\over z_2}\right)
+\delta\left({1\over z_1} -{1\over z_2}\right)\left( {-1\over 2}\DGhat_T^{(1)}(z_1)+{i\over 4}\DHhat_T^{(1)}(z_1)\right),
\label{FFDF1}\\
&&\Dhat_2\left({1\over z_1},{1\over z_2}\right)={\cal P}{-1\over {1\over z_2}-{1\over z_1}} 
\Nhat_2\left({1\over z_2}-{1\over z_1},{1\over z_2}\right)
+\delta\left({1\over z_1} -{1\over z_2}\right)\left( {i\over 2}\Ghat_T^{(1)}(z_1)-{i\over 4}
\DHhat_T^{(1)}(z_1)\right),
\label{FFDF2}\\
&&\Dhat_3\left({1\over z_1},{1\over z_2}\right)={\cal P}{1\over {1\over z_2}-{1\over z_1}} 
\Nhat_2\left({1\over z_1},{1\over z_2}\right)
+\delta\left({1\over z_1} -{1\over z_2}\right)
\left( {1\over 2}\DGhat_T^{(1)}(z_1)+{i\over 4}\DHhat_T^{(1)}(z_1)\right).  \qquad\quad
\label{FFDF3}
\eeq
These relations show $\Dhat_{1,2,3}\left({1\over z_1},{1\over z_2}\right)$ are 
completely determined by $\Nhat_{1,2}\left({1\over z_1},{1\over z_2}\right)$
and the kinematical FFs.  
Following the same procedure leading to (\ref{EOMrelation}) 
from (\ref{PDFDFrelation}), we can derive
the e.o.m. relation by contracting (\ref{FFDFrelation}) with $g^\perp_{\beta\gamma}$ as
\beq
&&{1\over z}\left( \DGhat_{\3Tb}(z)+i \DGhat_{3T}(z)\right)
-i\Dtilde_{FT}(z)
\nonumber\\
&&\qquad=i\int d\left({1\over z_1}\right)  P\left( {1\over {1\over z}- {1\over z_1}}\right)
\left\{ 
-2\Nhat_{1} \left( {1\over z_1},{1\over z}\right)- \Nhat_{2}\left({1\over z_1},{1\over z}\right) +\Nhat_{2}\left( {1\over z}-{1\over z_1},{1\over z}\right)
\right\}\nonumber\\
&&\qquad\qquad+{1\over 2}\left( \Ghat_T^{(1)}(z) + \Delta\Hhat_T^{(1)}(z)\right) +{i\over 2}\DGhat_T^{(1)}(z),
\label{FFDF1}
\eeq
where $\Dtilde_{FT}(z)$ is defined from the dynamical FFs in (\ref{FFtilde}) as
\beq
\Dtilde_{FT}(z)\equiv {2\over C_F}\int_0^{1/z}d\left({1\over z_1}\right)\Dtilde_{FT}\left({1\over z_1},{1\over z_1}-{1\over z}\right),
\label{DtildeFT}
\eeq
with $C_F={N^2-1\over 2N}$, 
and it has a support on $0<z<1$.  
Real and imaginary part of (\ref{FFDF1}), respectively, reads
\beq
&&{1\over z}\DGhat_{\3Tb}(z)
+\Im\Dtilde_{FT}(z)
\nonumber\\
&&\qquad=\int d\left({1\over z_1}\right)  P\left( {1\over {1\over z}- {1\over z_1}}\right)\Im
\left\{ 
2\Nhat_{1} \left( {1\over z_1},{1\over z}\right)+ \Nhat_{2}\left({1\over z_1},{1\over z}\right) -\Nhat_{2}\left( {1\over z}-{1\over z_1},{1\over z}\right)
\right\}\nonumber\\
&&\qquad+{1\over 2}\left( \Ghat_T^{(1)}(z) + \Delta\Hhat_T^{(1)}(z)\right), 
\label{FFDFodd1}
\eeq
and
\beq
&&{1\over z}\DGhat_{3T}(z)
-\Re\Dtilde_{FT}(z)
\nonumber\\
&&\qquad=\int d\left({1\over z_1}\right)  P\left( {1\over {1\over z}- {1\over z_1}}
\right)\Re
\left\{ 
-2\Nhat_{1} \left( {1\over z_1},{1\over z}\right)- \Nhat_{2}\left({1\over z_1},{1\over z}\right) +\Nhat_{2}\left( {1\over z}-{1\over z_1},{1\over z}\right)
\right\}\nonumber\\
&&\qquad\qquad+{1\over 2}\DGhat_T^{(1)}(z).  
\label{FFDFeven1}
\eeq
The relation (\ref{FFDFeven1}) is the FF version of (\ref{EOMrelation}).  

We can also derive another relation from (\ref{FFDFrelation}).  Following a similar step
from (\ref{PDFDF2}) to (\ref{DFrelation3}), we obtain the following relation. 
\beq
{\DHhat_{\3Tb}(z)\over z}+i {\DHhat_{3T}(z)\over z}=i\int d\left({1\over z_1}\right)
\left[ \Dhat_2\left({1\over z_1},{1\over z}\right) -\Dhat_3\left({1\over z_1},{1\over z}\right)\right].  
\eeq
Using (\ref{FFDF2}) and (\ref{FFDF3}) in the r.h.s. of this equation
and comparing real and imaginary parts of  both sides, 
one obtains the following two relations.   
\beq
&&{\DHhat_{\3Tb}(z)\over z}=
\int d\left({1\over z_1}\right){\cal P}{1\over {1\over z}-{1\over z_1}}
\Im\left[ \Nhat_2\left({1\over z_1},{1\over z}\right) +
\Nhat_2\left({1\over z}- {1\over z_1},{1\over z}\right)\right]\nonumber\\
&&\qquad\qquad \qquad+{1\over 2}\left( \DHhat_T^{(1)}(z)-\Ghat_T^{(1)}(z)\right),
\label{FFDFodd2}\\
&&{\DHhat_{3T}(z)\over z}=
\int d\left({1\over z_1}\right){\cal P}{-1\over {1\over z}-{1\over z_1}}
\Re\left[ \Nhat_2\left({1\over z_1},{1\over z}\right) +
\Nhat_2\left({1\over z}- {1\over z_1},{1\over z}\right)\right]\nonumber\\
&&\qquad\qquad \qquad-{1\over 2}\Delta\Ghat_T^{(1)}(z). 
\label{FFDFeven2}
\eeq
The second one is the FF version of (\ref{DFrelation2}) for the distribution function.  

To summarize this section, we have derived two independent relations
among the intrinsic, kinematical and dynamical functions, 
(\ref{FFDFeven1}) and (\ref{FFDFeven2}),  for the ``$T$-even'' sector,
and two independent ones (\ref{FFDFodd1}) and (\ref{FFDFodd2}) 
for the ``$T$-odd'' sector.  
One needs another independent relation for the former and two more relations
for the latter.  

\subsection{Constraint relations from nonlocal operator product expansion}

In this subsection, we will derive the relations among the twist-3 gluonic FFs, 
employing the method used in section 2.3.   
To this end, 
we consider operator identities for the correlation functions away from the lightcone 
which become the fragmentation matrix element in the lightlike limit.  
We need to calculate a matrix element like
\beq
{\partial \over \partial y_\rho}
\la 0|\left([\infty y,-y]F^{\alpha\nu}(-y)\right)_a|hX\ra\la hX|
\left( F^{\beta\mu}(y)[y,\infty y]\right)_a|0\ra, 
\label{FFcorrelation1}
\eeq
for $y^2\neq 0$ and take the $y^\mu\to\delta^{\mu}_- y^-$ limit after
differentiation.   
To calculate (\ref{FFcorrelation1}), we use the following operator identities:
\beq
&&\left(F^{\beta\mu}(y)[y,\infty y]\right)_a {\overleftarrow{\partial}\over \partial y_\rho}\nonumber\\
&&\qquad=\left(F^{\beta\mu}(y)\overleftarrow{D}^\rho (y) [y,\infty y]\right)_a
+i\int_{\infty}^1dt\,t \left(F^{\beta\mu}(y)[y,ty]gF^{\rho y}(ty)[ty,\infty y]\right)_a,
\label{FFderiv1}\\
&&{\partial \over \partial y_\rho}\left( [\infty y,-y]F^{\alpha\nu}(-y)\right)_a\nonumber\\
&&\quad
=-\left( [\infty y,-y]D^\rho(-y)F^{\alpha\nu}(-y)\right)_a
+i\int_{-1}^{\infty}dt\,t \left( [\infty y,ty]gF^{\rho y}(ty)[ty,-y]F^{\alpha\nu}(-y)\right)_a.\qquad
\label{FFderiv2}
\eeq
From translational invariance, we also have the relation
\beq
&&0=\lim_{\xi\to 0}{d\over d\xi_\rho}
\la 0|\left( [\infty y+\xi,-y+\xi]F^{\alpha\nu}(-y+\xi)\right)_a|hX\ra
\la hX| \left( F^{\beta\mu}(y+\xi)[y+\xi,\infty y+\xi]\right)_a|0\ra \nonumber\\
&&
=\la 0|\left([\infty y,-y]F^{\alpha\nu}(-y)\right)_a|hX\ra
\la hX|\left(F^{\beta\mu}(y)\overleftarrow{D}^\rho(y)[y,\infty y]\right)_a
|0\ra\nonumber\\
&&
+\la 0|\left([\infty y,-y]F^{\alpha\nu}(-y)\right)_a|hX\ra
\la hX| i\int_\infty^1dt \left( F^{\beta\mu}(y)[y,ty]gF^{\rho y}(ty)[ty,\infty y]\right)_a |0\ra 
\nonumber\\
&&+
\la 0|\left([\infty y,-y]D^\rho(-y)F^{\alpha\nu}(-y)\right)_a|hX\ra
\la hX| \left(F^{\beta\mu}(y)[y,\infty y]\right)_a|0\ra\nonumber\\
&&+
\la 0|i\int_{-1}^\infty dt \left( [\infty y,-y]
gF^{\rho y}(ty)[ty,-y]F^{\alpha\nu}(-y)\right)_a|hX\ra
\la hX| \left(F^{\beta\mu}(y)[y,\infty y]\right)_a|0\ra.\qquad
\label{translational}
\eeq
In (\ref{FFcorrelation1})-(\ref{translational}), we have explicitly written 
gauge links and color indices.  
Below we will suppress them for brevity.  
Inserting (\ref{FFderiv1}) and (\ref{FFderiv2}) into (\ref{FFcorrelation1}),
and using (\ref{translational}) to eliminate the term 
containing $\la 0|D^\rho(-y)F^{\alpha\nu}(-y)|hX\ra$, one obtains
\beq
&&{\partial \over \partial y_\rho}
\la 0|F^{\alpha\nu}(-y)|hX\ra\la hX|F^{\beta\mu}(y)|0\ra\nonumber\\
&&\qquad=
2\la 0|F^{\alpha\nu}(-y)|hX\ra\la hX|F^{\beta\mu}(y)\overleftarrow{D}^\rho(y)
|0\ra\nonumber\\
&&\qquad +\la 0|F^{\alpha\nu}(-y)|hX\ra\la hX|
i\int_\infty^1dt\,(t+1)F^{\beta\mu}(y) gF^{\rho y}(ty)|0\ra\nonumber\\
&&\qquad +\la 0|i\int^\infty_{-1}dt\,(t+1) gF^{\rho y}(ty) F^{\alpha\nu}(-y)|hX\ra\la hX|F^{\beta\mu}(y) |0\ra.  
\label{FFcorrelation2}
\eeq
This equation is the starting point of our analysis in this section. 
Constraint relations for the twist-3 FFs can be obtained by expressing each term
of (\ref{FFcorrelation2}) in terms of the FFs defined in Sec. 3.1.  
To calculate the l.h.s. of  (\ref{FFcorrelation2}), we need the
Fourier inversion of (\ref{FF1}) 
for non-lightlike separation ($y^2\neq 0$), which can be written as
\beq
&&\hspace{-0.3cm} {1\over N^2-1} \sum_X \la 0| F^{\alpha\nu}(-y) |hX\ra\la hX|
F^{\beta\mu}(y) |0\ra\nonumber\\
&&=\int d\left({1\over z}\right)e^{2iP_h\cdot y/z}
\Biggl[
\left\{ 
\left( -g_\perp^{\alpha\beta}P_h^\mu+g_\perp^{\alpha\mu}P_h^\beta \right)P_h^\nu -
\left( -g_\perp^{\nu\beta}P_h^\mu+g_\perp^{\nu\mu}P_h^\beta \right)P_h^\alpha 
\right\}\Ghat(z)
\Biggr.
\nonumber\\
&&\left. +{i(S\cdot y) \over (P_h\cdot y)^2}
\left\{ \left( 
\epsilon^{P_h y \alpha \beta}P_h^\mu -  \epsilon^{P_h y \alpha \mu}P_h^\beta \right)P_h^\nu 
-
\left( 
\epsilon^{P_h y \nu \beta}P_h^\mu -  \epsilon^{P_h y \nu \mu}P_h^\beta \right)P_h^\alpha 
\right\} \DGhat(z)\right.\nonumber\\
&&\left.
- {M_h \over (P_h\cdot y)^2}  \left[
\left\{ 
\left(y^\alpha\epsilon^{\beta P_h y S_\perp } + y^\beta\epsilon^{\alpha P_h y S_\perp } \right) P_h^\mu 
-
\left(y^\alpha\epsilon^{\mu P_h y S_\perp } + y^\mu \epsilon^{\alpha P_h y S_\perp } \right) P_h^\beta
\right\} P_h^\nu\right.\right.\nonumber\\
&&\left.
\left. \quad\ \ -
\left\{ 
\left(y^\nu\epsilon^{\beta P_h y S_\perp } + y^\beta\epsilon^{\nu P_h y S_\perp } \right) P_h^\mu 
-
\left(y^\nu\epsilon^{\mu P_h y S_\perp } + y^\mu \epsilon^{\nu P_h y S_\perp } \right) P_h^\beta
\right\} P_h^\alpha\
\right]\DGhat_{\3Tb} (z)\right.\nonumber\\
&&\left.- { iM_h \over P_h\cdot y} 
\left\{ 
\left( \epsilon^{\alpha \beta y S_\perp}P_h^\mu -\epsilon^{\alpha\mu y S_\perp} P_h^\beta\right) P_h^\nu
-
\left( \epsilon^{\nu \beta y S_\perp}P_h^\mu -\epsilon^{\nu \mu y S_\perp} P_h^\beta\right) P_h^\alpha
\right\} \DGhat_{3T}(z)\right.
\nonumber\\
&&\left.
+{M_h \over P_h\cdot y} \left\{ 
\epsilon^{\beta\mu P_h y}
\Bigl( S_\perp^\alpha P_h^\nu - S_\perp^\nu P_h^\alpha \Bigr)+
\epsilon^{\alpha\nu P_h y}\left( S_\perp^\beta P_h^\mu - S_\perp^\mu P_h^\beta \right)
\right\} \DHhat_{\3Tb}(z)\right.
\nonumber\\
&&
\Biggl.
+  {i M_h \over P_h\cdot y}  \left\{ 
\epsilon^{\beta\mu P_h y}\Bigl( S_\perp^\alpha P_h^\nu - S_\perp^\nu P_h^\alpha \Bigr)
-
\epsilon^{\alpha\nu P_h y}\left( S_\perp^\beta P_h^\mu - S_\perp^\mu P_h^\beta \right)
\right\} \DHhat_{3T}(z)+\cdots
\Biggr].  
\label{FF2}
\eeq
In calculating the derivative of the l.h.s. of  (\ref{FFcorrelation2}), 
one need to use $S_\perp^\mu = S^\mu -{S\cdot y\over P_h\cdot y}P_h^\mu$
and $g_\perp^{\mu\nu}=g^{\mu\nu}- {1\over P_h\cdot y}(P_h^\mu y^\nu + P_h^\nu y^\mu)$ in (\ref{FF2}).  
This way the l.h.s. of (\ref{FFcorrelation2}) can be written in terms of the
intrinsic FFs in (\ref{FF2}).  
Likewise the second and the third terms in the r.h.s. of (\ref{FFcorrelation2}) can be easily
expressed by using the dynamical FFs in (\ref{FFFtype}).  In order to express
the first term in the r.h.s. of (\ref{FFcorrelation2}) in terms of the dynamical FFs, 
we introduce two particular contractions with respect to the Lorentz indices 
which allows use of the QCD e.o.m. 
$ F_a^{\mu\alpha}(y) \overleftarrow{D}_\mu (y)=-g\bar{\psi}(y)\gamma^\alpha t^a \psi(y)$.

\subsubsection{Relations from operator identity I}
We can obtain a constraint relation from the following identity:  
\beq
&&\hspace{-0.5cm}
y_\rho \left[ {\partial \over \partial y_\rho} \la 0 | F^{\mu y}(-y) | hX \ra \la hX | F_\mu^{\ \alpha}(y) | 0 \ra 
- ( \alpha \leftrightarrow \rho) \right]\nonumber\\
&&= \la 0 | F^{\mu y}(-y) | hX \ra \la hX | F_\mu^{\ \alpha }(y) | 0 \ra - 
\la 0 | F^{\mu \alpha}(-y) | hX \ra \la hX | F_\mu^{\ y}(y) | 0 \ra\nonumber\\
&& +2 \la 0 | g\bar{\psi}(-y)\yslash t^a \psi(-y) | hX \ra \la hX | F_{a}^{\ y \alpha}(y) | 0 \ra
\nonumber\\
&&-2 if^{abc}\la 0 | F_a^{\ \mu y}(-y) | hX \ra \la hX | ig\int_\infty^1dt F_b^{\ y\alpha}(y)F_{c\mu}^{\ \ y}(ty) | 0 \ra\nonumber\\
&&- if^{abc}\la 0 | F_a^{\ \mu y}(-y) | hX \ra \la hX | ig\int_\infty^1dt (t+1) F_{b\,\mu}^{\ \ y}(y)F_{c}^{\ \alpha y}(ty) | 0 \ra\nonumber\\
&&-2 if^{abc}\la 0 | ig \int_{-1}^\infty dt F_a^{\ \mu y}(ty) F_{b\mu}^{\ \ y}(-y) | hX \ra \la hX | F_{c}^{\  y\alpha}(y) | 0 \ra\nonumber\\
&&- if^{abc}\la 0 | ig \int_{-1}^\infty dt (t+1) F_a^{\alpha  y}(ty) F_{b\mu}^{\ \ y}(-y) | hX \ra \la hX | F_{c}^{\mu y}(y) | 0 \ra. 
\label{FFidentity1}
\eeq
This identity can be obtained as follows:
We first use (\ref{FFcorrelation2}) in the l.h.s. of (\ref{FFidentity1}).  We then find that
the terms corresponding to the first term in the r.h.s. of (\ref{FFcorrelation2})
read
$$\la 0 | F^{\mu y}(-y) | hX \ra\la hX| \left(F_\mu^{\ \alpha} (y)
\overleftarrow{D}^\rho (y) y_\rho- 
F_\mu^{\ \rho} (y)\overleftarrow{D}^\alpha (y) y_\rho \right)|0\ra,$$
which is equal to
$$-\la 0 | F^{\mu y}(-y) | hX \ra\la hX| F^{\alpha\rho} (y)
\overleftarrow{D}_\mu (y) y_\rho|0\ra,$$ 
by the Bianchi identity.  
Then by using the relation (\ref{translational}),
it is transformed into
$$\la 0 | y_\beta D_\mu (y) F^{\mu \beta}(-y) | hX \ra\la hX| F^{\alpha\rho} (y) 
y_\rho|0\ra,$$ plus terms which contain three field strengths.
The former eventually becomes the third term in the r.h.s. of (\ref{FFidentity1})
by the QCD e.o.m., and the latter is shown as the fourth and the sixth terms 
in the r.h.s. of (\ref{FFidentity1}).  

Using (\ref{FF2}) and the inverse Fourier transform of (\ref{FFFtype}) in (\ref{FFidentity1}), 
one obtains the following relation among the twist-3 fragmentation functions:
\beq
&&
-{\partial \over \partial (1/z)}\left\{ 
{1\over z}\left( \DGhat_{\3Tb}(z) -i \DGhat_{3T}(z)\right)\right\} +2\DGhat_{\3Tb}(z) \nonumber\\
&&\qquad\qquad+{\partial \over \partial (1/z)}\left\{ 
{1\over z}\left( \DHhat_{\3Tb}(z) + i \DHhat_{3T}(z)\right)\right\}
- 2\DHhat_{\3Tb}(z) \nonumber\\
&&\qquad
=
i{\partial\over \partial (1/z)}\Dtilde_{FT}^*(z)
+i \int d\left({1\over z'}\right){1\over {1\over z}-{1\over z'} +i\epsilon}
{\partial \over \partial (1/z)}\left\{
-\Nhat_{2}\left({1\over z'},{1\over z}\right) -\Nhat_{2}\left( {1\over z}-{1\over z'},{1\over z}\right)
\right\}
\nonumber\\
&&\qquad+i \int d\left({1\over z'}\right) {1\over \left({1\over z}-{1\over z'}+i\epsilon\right)^2}
\left\{ 
\Nhat_1 \left( {1\over z'},{1\over z}\right)+2\Nhat_{2}\left({1\over z'},{1\over z}\right) -\Nhat_{2}\left( {1\over z}-{1\over z'},{1\over z}\right)
\right\}
\nonumber\\
&&\qquad+i \int d\left({1\over z'}\right){1\over {1\over z}-{1\over z'} -i\epsilon}
{\partial \over \partial (1/z)}\left\{
-2\Nhat_{1}^* \left({1\over z'},{1\over z}\right) 
-\Nhat_{2}^* \left({1\over z'},{1\over z}\right) +\Nhat_{2}^*\left( {1\over z}-{1\over z'},{1\over z}\right)
\right\}
\nonumber\\
&&\qquad+i \int d\left({1\over z'}\right) {1\over \left({1\over z}-{1\over z'}-i\epsilon\right)^2}
\left\{ 
\Nhat_1^* \left( {1\over z'},{1\over z}\right)+\Nhat_{2}^*\left( {1\over z}-{1\over z'},{1\over z}\right)
\right\}. 
\label{FFLIR1}
\eeq
The real part of this equation reads
\beq
&&
-{\partial \over \partial (1/z)}\left\{ 
{1\over z}\left( \DGhat_{\3Tb}(z) - \DHhat_{\3Tb}(z)\right)\right\} 
+2\left( \DGhat_{\3Tb}(z) - \DHhat_{\3Tb}(z)\right) \nonumber\\
&&\qquad=
{\partial\over \partial (1/z)}\Im\Dtilde_{FT}(z)
+ \int d\left({1\over z'}\right){1\over {1\over z}-{1\over z'} }
{\partial \over \partial (1/z)}\Im
\left\{
-2\Nhat_{1}\left({1\over z'},{1\over z}\right) 
+2 \Nhat_{2}\left( {1\over z}-{1\over z'},{1\over z}\right)
\right\}
\nonumber\\
&&\qquad+ \int d\left({1\over z'}\right) {1\over \left({1\over z}-{1\over z'}\right)^2} \Im 
\left\{ 
- 2\Nhat_{2}\left({1\over z'},{1\over z}\right) +2\Nhat_{2}\left( {1\over z}-{1\over z'},{1\over z}\right)
\right\}, 
\label{FFLIRodd1}
\eeq
and the imaginary part gives 
\beq
&&
{\partial \over \partial (1/z)}\left\{ 
{1\over z}\left( \DGhat_{3T}(z) + \DHhat_{3T}(z)\right)\right\}  \nonumber\\
&&\qquad=
{\partial\over \partial (1/z)}\Re\Dtilde_{FT}(z)
+ \int d\left({1\over z'}\right){1\over {1\over z}-{1\over z'} }
{\partial \over \partial (1/z)}\Re
\left\{
-2\Nhat_{1}\left({1\over z'},{1\over z}\right) - 2 \Nhat_{2}\left( {1\over z'},{1\over z}\right)
\right\}
\nonumber\\
&&\qquad+ \int d\left({1\over z'}\right) {1\over \left({1\over z}-{1\over z'}\right)^2} \Re
\left\{ 
 2\Nhat_{1}\left({1\over z'},{1\over z}\right) +2\Nhat_{2}\left( {1\over z'},{1\over z}\right)
\right\}.  
\label{FFLIReven1}
\eeq
Equations (\ref{FFLIRodd1}) and (\ref{FFLIReven1}) are
the constraint relations among the intrinsic and the dynamical FFs.
We note that (\ref{FFLIReven1}) is the relation obtained as the sum of
(\ref{FFDFeven1}) and (\ref{FFDFeven2}), while (\ref{FFLIRodd1}) is
an independent relation from (\ref{FFDFodd1}) and (\ref{FFDFodd2}).

\subsubsection{Relations from operator identity II}
Here we use the following identity to get independent relations among the twist-3 FFs:
\beq
&&{\partial \over \partial y^\mu} \la 0 | F^{y\nu}(-y) |hX\ra \la hX|F^{y\mu}(y)|0\ra\nonumber\\
&&\qquad=\la 0 | F_\mu^{\ \nu}(-y) |hX\ra \la 0|F^{y\mu}(y)|0\ra
+2 \la 0 | F^{y \nu}(-y) |hX\ra \la 0| g\bar{\psi}(y)\yslash t^a \psi(y) |0\ra\nonumber\\
&&\qquad + if^{abc} \la 0 | F_a^{\ y \nu}(-y) |hX\ra \la 0| ig\int_\infty^1 dt (t+1)F_b^{\ y\rho}(y)F_{c\,\rho}^{\ \ y}(ty)|0\ra
\nonumber\\
&&\qquad+ if^{abc} \la 0 | ig \int_{-1}^\infty dt(t+1) F_{a\rho}^{\ \ y}(ty)F_b^{\ y\nu}(-y) |hX\ra \la 0|F_{c}^{\ y\rho}(ty)|0\ra.  
\label{FFcorrelator2}
\eeq
This relation is obtained by using (\ref{FFcorrelation2}) in the l.h.s.of (\ref{FFcorrelator2}),
and taking into account of the QCD e.o.m., 
$D_\mu (y) F_a^{\mu\alpha}(y)  =-g\bar{\psi}(y)\gamma^\alpha t^a\psi(y)$.  
For the calculation of the l.h.s. of this equation, one should 
use (\ref{FF2}) contracted with 
$y^\alpha y^\beta$ by keeping $y^2\neq 0$ before taking the derivative:
\beq
&&{1\over N^2-1}\sum_X  \la 0 | F^{y\nu}(-y) |hX\ra \la hX|F^{y\mu}(y)|0\ra\nonumber\\
&&=\int d\left({1\over z}\right) e^{2iP_h\cdot y/z}
\left[
\left\{
-g_\perp^{\alpha\beta}y_\alpha y_\beta P_h^\mu P_h^\nu +(P_h\cdot y)
\left( g_\perp^{\mu y} P_h^\nu + g_\perp^{\nu y} P_h^\mu \right) -g_\perp^{\nu\mu}(P_h\cdot y)^2\right\}\Ghat(z)
\right.\nonumber\\
&&\left.\qquad\qquad +i(S\cdot y)\epsilon^{P_hy\nu\mu}\DGhat(z)
-iM_h\epsilon^{\nu\mu y S_\perp} (P_h\cdot y)\DGhat_{3T}(z)
\right.\nonumber\\
&&\left.\qquad
+M_h\left\{ 
{y^2\over P_h\cdot y}\left( \epsilon^{\mu P_h y S_\perp} P_h^\nu + \epsilon^{\nu P_h y S_\perp} P_h^\mu \right)
- \left( \epsilon^{\mu P_h y S_\perp} y^\nu + \epsilon^{\nu P_h y S_\perp} y^\mu \right)
\right\}\DGhat_{\3Tb}(z)
\right]. 
\label{FFcorrelator3}
\eeq
From the identity (\ref{FFcorrelator2}), one obtains the following relation:
\beq
&&
{\partial \over \partial (1/z)}\left\{ 
{1\over z} \DGhat_{\3Tb}(z) \right\}
+
{i\over z}{\partial \over \partial (1/z)}\DGhat_{3T}(z)
-3\DGhat_{\3Tb}(z) - \DHhat_{\3Tb}(z) +i\DHhat_{3T}(z) + 
i \Delta \widehat{G}(z) 
\nonumber\\
&&=
i{\partial\over \partial (1/z)}\Dtilde_{FT}(z)
\nonumber\\
&&-i \int d\left({1\over z'}\right){1\over {1\over z}-{1\over z'} +i\epsilon}
{\partial \over \partial (1/z)}\left\{
2\Nhat_{1}\left( {1\over z'},{1\over z}\right)+\Nhat_{2}\left({1\over z'},{1\over z}\right) -\Nhat_{2}\left( {1\over z}-{1\over z'},{1\over z}\right)
\right\}
\nonumber\\
&&+i \int d\left({1\over z'}\right) {1\over \left({1\over z}-{1\over z'}-i\epsilon\right)^2}
\left\{ 
\Nhat_{1}^* \left( {1\over z'},{1\over z}\right)
+2\Nhat_{2}^*\left({1\over z'},{1\over z}\right) 
-\Nhat_{2}^*\left( {1\over z}-{1\over z'},{1\over z}\right)
\right\}.  
\label{corr2relation}
\eeq
The real part of this equation gives
\beq
&&{\partial \over \partial (1/z)}\left\{ 
{1\over z} \DGhat_{\3Tb}(z) \right\} -3\DGhat_{\3Tb}(z) - \DHhat_{\3Tb}(z) \nonumber\\
&&\qquad=
-{\partial\over \partial (1/z)}\Im\Dtilde_{FT}(z)
\nonumber\\
&&\qquad
+\int d\left({1\over z'}\right){1\over {1\over z}-{1\over z'} }
{\partial \over \partial (1/z)}\Im\left\{
2\Nhat_{1}\left( {1\over z'},{1\over z}\right)
+\Nhat_{2}\left({1\over z'},{1\over z}\right) 
-\Nhat_{2}\left( {1\over z}-{1\over z'},{1\over z}\right)
\right\}
\nonumber\\
&&\qquad+ \int d\left({1\over z'}\right) {1\over \left({1\over z}-{1\over z'}\right)^2}\Im
\left\{ 
\Nhat_{1} \left( {1\over z'},{1\over z}\right)
+2\Nhat_{2}\left({1\over z'},{1\over z}\right) 
-\Nhat_{2}\left( {1\over z}-{1\over z'},{1\over z}\right)
\right\}, 
\label{FFLIRodd2}
\eeq
while the imaginary part is
\beq
&&
{1\over z} {\partial \over \partial (1/z)}
\left\{ 
\DGhat_{3T}(z) 
\right\} 
+ \DHhat_{3T}(z) +
\Delta \widehat{G}(z)
\nonumber\\
&&=
{\partial\over \partial (1/z)}\Re\Dtilde_{FT}(z)
-\int d\left({1\over z'}\right){1\over {1\over z}-{1\over z'} }
{\partial \over \partial (1/z)}\Re \left\{
2\Nhat_{1}\left( {1\over z'},{1\over z}\right)+
\Nhat_{2}\left({1\over z'},{1\over z}\right) 
-\Nhat_{2}\left( {1\over z}-{1\over z'},{1\over z}\right)
\right\}
\nonumber\\
&&+ \int d\left({1\over z'}\right) {1\over \left({1\over z}-{1\over z'}\right)^2}\Re
\left\{ 
\Nhat_{1} \left( {1\over z'},{1\over z}\right)+2\Nhat_{2}\left({1\over z'},{1\over z}\right) -\Nhat_{2}\left( {1\over z}-{1\over z'},{1\over z}\right)
\right\}. 
\label{FFLIReven2}
\eeq

For consistency check, we have also analyzed the correlation function
\beq
y_\rho \left[ {\partial \over \partial y_\rho} \la 0 | F^{\mu y}(-y) | hX \ra \la hX | \widetilde{F}_\mu^{\ \alpha}(y) | 0 \ra 
- ( \alpha \leftrightarrow \rho) \right].
\eeq
This operator only gives the relation among the $T$-even  functions which is identical to (\ref{FFLIReven2}).

To summarize this section, from nonlocal OPE we have derived 
an independent relation (\ref{FFLIReven2}) for the $T$-even sector,
and two independent relations (\ref{FFLIRodd1}) and (\ref{FFLIRodd2})
for the $T$-odd sector.

\subsection{Solution for intrinsic and kinematical FFs in terms of dynamical FFs}

Using the constraint relations derived 
in subsections 3.2 and 3.3, we present here
expressions for the intrinsic and the kinematical FFs
in terms of the twist-2 FFs and the twist-3 dynamical FFs.  
Since
$\Nhat_1\left({1\over z_1},{1\over z_2}\right)$,  $\Nhat_2\left({1\over z_1},{1\over z_2}\right) $ and 
$\Nhat_2\left({1\over z_2}-{1\over z_1},{1\over z_2}\right) $ 
have a support
on ${1\over z_2} > 1$ and ${1\over z_2} > {1\over z_1} >0$,
they vanish at the edge of the support,
i.e., $\Nhat_{1,2}\left( 0,{1\over z}  \right)=\Nhat_{1,2}\left( {1\over z},{1\over z}  \right)
=\Nhat_{1,2}\left( {1\over z},1 \right)=0$.
$\Dtilde_{FT}(z)$ has a support on $z<1$ and thus $\Dtilde_{FT}(1)=0$.  
Taking these boundary conditions into account, we can integrate the constraint relations.  

\subsubsection{$T$-odd fragmentation functions}
We first integrate (\ref{FFLIRodd1}) to obtain $\DGhat_{\3Tb}(z)-\DHhat_{\3Tb}(z)$.
From that result and (\ref{FFLIRodd2}) one obtains
$\DHhat_{\3Tb}(z)$ and $\DGhat_{\3Tb}(z)$ in terms of 
the dynamical FFs.  
Since the calculation is straightforward, we only present the final result.  
The result for $\DGhat_{\3Tb}(z)$ reads
\beq
&&\DGhat_{\3Tb}(z)=-z\,\Im\Dtilde_{FT}(z)-{2\over z^3}\int_1^{1/z}d\left({1\over z_2}\right)z_2^5\,\Im \Dtilde_{FT}(z_2)
-{1\over z}\int_1^{1/z}d\left({1\over z_2}\right)z_2^3\,\Im\Dtilde_{FT}(z_2)\nonumber\\
&&\quad
+z\int_0^{1/z}d\left({1\over z_1}\right){1\over 1/z-1/z_1}\Im\left[ 
2\Nhat_1\left({1\over z_1},{1\over z}\right) + \Nhat_2\left({1\over z_1},{1\over z}\right) -\Nhat_2\left(  {1\over z}- {1\over z_1} ,{1\over z}\right)
\right]\nonumber\\
&&\quad+{4\over z^3}\int_1^{1/z}d\left({1\over z_2}\right)z_2^5\int_0^{1/z_2}d\left({1\over z_1}\right){1\over 1/z_2-1/z_1}\Im\left[ 
\Nhat_1\left({1\over z_1},{1\over z_2}\right) + \Nhat_2\left({1\over z_1},{1\over z_2}\right) 
\right]\nonumber\\
&&\quad+{2\over z^3}\int_1^{1/z}d\left({1\over z_2}\right)z_2^4 \int_0^{1/z_2}d\left({1\over z_1}\right){1\over \left(1/z_2-1/z_1\right)^2}\Im\left[ 
\Nhat_1\left({1\over z_1},{1\over z_2}\right) + \Nhat_2\left({1\over z_1},{1\over z_2}\right) 
\right]\nonumber\\
&&\quad+{2\over z}\int_1^{1/z}d\left({1\over z_2}\right)z_2^3\int_0^{1/z_2}d\left({1\over z_1}\right){1\over 1/z_2-1/z_1}\Im\left[ 
\Nhat_1\left({1\over z_1},{1\over z_2}\right) - \Nhat_2\left({1\over z_2}-{1\over z_1},{1\over z_2}\right) 
\right]\nonumber\\
&&\quad+{1\over z}\int_1^{1/z}d\left({1\over z_2}\right)z_2^2\int_0^{1/z_2}d\left({1\over z_1}\right){1\over \left(1/z_2-1/z_1\right)^2}\Im\left[ 
\Nhat_1\left({1\over z_1},{1\over z_2}\right) +\Nhat_2\left({1\over z_1},{1\over z_2}\right) \right.\nonumber\\
&&\left.\hspace{9cm}- 2\Nhat_2\left({1\over z_2}-{1\over z_1},{1\over z_2}\right) 
\right].
\label{DG3Tb}
\eeq
Integrals in this equation can be rewritten as 
\beq
\int_1^{1/z}d\left({1\over z_2}\right) \int_0^{1/z_2}d\left({1\over z_1}\right)\cdots =
\int^1_{z}{dz_2\over z_2^2} \int^\infty_{z_2}{dz_1\over z_1^2}\cdots.
\eeq
Similarly the result for $\DHhat_{\3Tb}(z)$ is given by
\beq
&&\DHhat_{\3Tb}(z)=-{2\over z^3}\int_1^{1/z}d\left({1\over z_2}\right)z_2^5\,\Im \Dtilde_{FT}(z_2)
+{1\over z}\int_1^{1/z}d\left({1\over z_2}\right)z_2^3\,\Im\Dtilde_{FT}(z_2)\nonumber\\
&&\quad
+z\int_0^{1/z}d\left({1\over z_1}\right){1\over 1/z-1/z_1}\Im\left[ 
\Nhat_2\left({1\over z_1},{1\over z}\right) +\Nhat_2\left(  {1\over z}- {1\over z_1} ,{1\over z}\right)
\right]\nonumber\\
&&\quad+{4\over z^3}\int_1^{1/z}d\left({1\over z_2}\right)z_2^5\int_0^{1/z_2}d\left({1\over z_1}\right){1\over 1/z_2-1/z_1}\Im\left[ 
\Nhat_1\left({1\over z_1},{1\over z_2}\right) + \Nhat_2\left({1\over z_1},{1\over z_2}\right) 
\right]\nonumber\\
&&\quad+{2\over z^3}\int_1^{1/z}d\left({1\over z_2}\right)z_2^4 \int_0^{1/z_2}d\left({1\over z_1}\right){1\over \left(1/z_2-1/z_1\right)^2}\Im\left[ 
\Nhat_1\left({1\over z_1},{1\over z_2}\right) + \Nhat_2\left({1\over z_1},{1\over z_2}\right) 
\right]\nonumber\\
&&\quad-{2\over z}\int_1^{1/z}d\left({1\over z_2}\right)z_2^3\int_0^{1/z_2}d\left({1\over z_1}\right){1\over 1/z_2-1/z_1}\Im\left[ 
\Nhat_1\left({1\over z_1},{1\over z_2}\right) - \Nhat_2\left({1\over z_2}-{1\over z_1},{1\over z_2}\right) 
\right]\nonumber\\
&&\quad-{1\over z}\int_1^{1/z}d\left({1\over z_2}\right)z_2^2\int_0^{1/z_2}d\left({1\over z_1}\right){1\over \left(1/z_2-1/z_1\right)^2}\Im\left[ 
\Nhat_1\left({1\over z_1},{1\over z_2}\right) +\Nhat_2\left({1\over z_1},{1\over z_2}\right) \right.\nonumber\\
&&\left.\hspace{9cm}- 2\Nhat_2\left({1\over z_2}-{1\over z_1},{1\over z_2}\right) 
\right].
\label{DH3Tb}
\eeq
Using (\ref{DG3Tb}) and (\ref{DH3Tb}) in (\ref{FFDFodd1}) and (\ref{FFDFodd2}), 
one can obtain the kinematical FFs as
\beq
&&\widehat{G}_T^{(1)}(z)=
-{2\over z^2}\int_1^{1/z}d\left({1\over z_2}\right)z_2^3\,\Im \Dtilde_{FT}(z_2)\nonumber\\
&&\quad+{4\over z^2}\int_1^{1/z}d\left({1\over z_2}\right)z_2^3\int_0^{1/z_2}d\left({1\over z_1}\right){1\over 1/z_2-1/z_1}\Im\left[ 
\Nhat_1\left({1\over z_1},{1\over z_2}\right) - \Nhat_2\left( {1\over z_2}-{1\over z_1} ,{1\over z_2}\right) 
\right]\nonumber\\
&&\quad+{2\over z^2}\int_1^{1/z}d\left({1\over z_2}\right)z_2^2 \int_0^{1/z_2}d\left({1\over z_1}\right){1\over \left(1/z_2-1/z_1\right)^2}\Im\left[ 
\Nhat_1\left({1\over z_1},{1\over z_2}\right) + \Nhat_2\left({1\over z_1},{1\over z_2}\right) \right.\nonumber\\
&&\left.\hspace{9cm}-2  \Nhat_2\left( {1\over z_2}-{1\over z_1} ,{1\over z_2}\right) 
\right],
\eeq
and
\beq
&&\Delta\widehat{H}_T^{(1)}(z)=
-{4\over z^4}\int_1^{1/z}d\left({1\over z_2}\right)z_2^5\,\Im \Dtilde_{FT}(z_2)\nonumber\\
&&\quad+{8\over z^4}\int_1^{1/z}d\left({1\over z_2}\right)z_2^5\int_0^{1/z_2}d\left({1\over z_1}\right){1\over 1/z_2-1/z_1}\Im\left[ 
\Nhat_1\left({1\over z_1},{1\over z_2}\right) + \Nhat_2\left({1\over z_1},{1\over z_2}\right) 
\right]\nonumber\\
&&\quad+{4\over z^4}\int_1^{1/z}d\left({1\over z_2}\right)z_2^4 \int_0^{1/z_2}d\left({1\over z_1}\right){1\over \left(1/z_2-1/z_1\right)^2}\Im\left[ 
\Nhat_1\left({1\over z_1},{1\over z_2}\right) + \Nhat_2\left({1\over z_1},{1\over z_2}\right) 
\right].
\eeq

\subsubsection{$T$-even fragmentation function}
The solution for $\DGhat_{3T}(z)$, $\DGhat_T^{(1)}(z)$ and $\DHhat_{3T}(z)$
can be obtained by integrating the relations
(\ref{FFDFeven1}), (\ref{FFDFeven2}) and (\ref{FFLIReven2}).  
Actually we can make a short cut.  Since they are
in parallel with 
(\ref{EOMrelation}), (\ref{DFrelation2}) and (\ref{OPE1})
for the gluon distributions $\Delta G_{3T}(x)$,
$\Delta G_T^{(1)}(x)$ and $\Delta H_{3T}(x)$,
we can read off the desired results from
(\ref{DelG3T}), (\ref{GT1}) and (\ref{DelH3T}) by a simple replacement.  
The results read
\beq
&&\DGhat_{3T}(z)=-{1\over z}\int_1^{1/z}d\left({1\over z_2}\right)z_2^2\DGhat(z_2)
+z\,\Re\Dtilde_{FT}(z)+{1\over z}\int_1^{1/z}d\left({1\over z_2}\right)z_2^3\,
\Re\Dtilde_{FT}(z_2)\nonumber\\
&&\qquad-z\int_0^{1/z}d\left({1\over z_1}\right){1\over {1\over z}-{1\over z_1}}
\Re\left[
2\Nhat_1\left( {1\over z_1},{1\over z}\right)
+\Nhat_2\left( {1\over z_1},{1\over z}\right)
-\Nhat_2\left( {1\over z} -{1\over z_1},{1\over z}\right)\right]\nonumber\\
&&\qquad-{2\over z}
\int_1^{1/z}d\left({1\over z_2}\right)z_2^3\int_0^{1/z_2}
d\left({1\over z_1}\right){1\over {1\over z_2}-{1\over z_1}}
\Re\left[
\Nhat_1\left( {1\over z_1},{1\over z_2}\right)
-\Nhat_2\left( {1\over z_2} -{1\over z_1},{1\over z_2}\right)\right]\nonumber\\
&&\qquad-{1\over z}
\int_1^{1/z}d\left({1\over z_2}\right)z_2^2\int_0^{1/z_2}
d\left({1\over z_1}\right){1\over \left({1\over z_2}-{1\over z_1}\right)^2}
\Re\left[
\Nhat_1\left( {1\over z_1},{1\over z_2}\right)
-\Nhat_2\left( {1\over z_1},{1\over z_2}\right)\right], 
\eeq
\beq
&&\DGhat_{T}^{(1)}(z)={1\over z^2}\int_1^{1/z}d\left({1\over z_2}\right)z_2^2\DGhat(z_2)
-{2\over z^2}\int_1^{1/z}d\left({1\over z_2}\right)z_2^3\,
\Re\Dtilde_{FT}(z_2)\nonumber\\
&&\qquad+{4\over z^2}
\int_1^{1/z}d\left({1\over z_2}\right)z_2^3\int_0^{1/z_2}
d\left({1\over z_1}\right){1\over {1\over z_2}-{1\over z_1}}
\Re\left[
\Nhat_1\left( {1\over z_1},{1\over z_2}\right)
-\Nhat_2\left( {1\over z_2} -{1\over z_1},{1\over z_2}\right)\right]\nonumber\\
&&\qquad+{2\over z^2}
\int_1^{1/z}d\left({1\over z_2}\right)z_2^2\int_0^{1/z_2}
d\left({1\over z_1}\right){1\over \left({1\over z_2}-{1\over z_1}\right)^2}
\Re\left[
\Nhat_1\left( {1\over z_1},{1\over z_2}\right)
-\Nhat_2\left( {1\over z_1},{1\over z_2}\right)\right], 
\eeq
and
\beq
&&\DHhat_{3T}(z)={1\over z}\int_1^{1/z}d\left({1\over z_2}\right)z_2^2\DGhat(z_2)
-{1\over z}\int_1^{1/z}d\left({1\over z_2}\right)z_2^3\,
\Re\Dtilde_{FT}(z_2)\nonumber\\
&&\qquad-z\int_0^{1/z}d\left({1\over z_1}\right){1\over {1\over z}-{1\over z_1}}
\Re\left[
\Nhat_2\left( {1\over z_1},{1\over z}\right)
+\Nhat_2\left( {1\over z} -{1\over z_1},{1\over z}\right)\right]\nonumber\\
&&\qquad+{2\over z}
\int_1^{1/z}d\left({1\over z_2}\right)z_2^3\int_0^{1/z_2}
d\left({1\over z_1}\right){1\over {1\over z_2}-{1\over z_1}}
\Re\left[
\Nhat_1\left( {1\over z_1},{1\over z_2}\right)
-\Nhat_2\left( {1\over z_2} -{1\over z_1},{1\over z_2}\right)\right]\nonumber\\
&&\qquad+{1\over z}
\int_1^{1/z}d\left({1\over z_2}\right)z_2^2\int_0^{1/z_2}
d\left({1\over z_1}\right){1\over \left({1\over z_2}-{1\over z_1}\right)^2}
\Re\left[
\Nhat_1\left( {1\over z_1},{1\over z_2}\right)
-\Nhat_2\left( {1\over z_1},{1\over z_2}\right)\right].  
\eeq
This completes the derivation of all the relations among the twist-3 gluonic FFs.  

To summarize this section,
we have derived all the constraint relations for twist-3 gluonic FFs,
which follow from the QCD e.o.m. and the operator product expansion.    
These relations are exact and need to be taken into
account in deriving a twist-3 cross section to which they contribute, and should 
constitute a cornerstone for proving the gauge invariance and Lorentz invariance of the
cross sections.
In particular, the intrinsic and kinematical
twist-3 FFs
are completely determined by the twist-2 FF and the dynamical twist-3
FFs (=three-gluon correlation functions), which provides a basis
for the renormalization of the intrinsic and kinematical FFs.

\section{Summary}

In this paper, we have performed a systematic study on the collinear twist-3 gluonic distribution 
functions (DFs) and fragmentation
functions (FFs).  
Both DFs and FFs are classified into three categories, intrinsic, kinematical and dynamical functions.  
Although they are convenient tools to describe twist-3 cross sections, they are not independent
of each other, but are constrained by a set of exact relations which follow from the QCD e.o.m. and
the nonlocal operator product expansion.   We have derived all those constraint relations
for all the gluonic twist-3 DFs and FFs and have given expressions for the intrinsic and kinematical
DFs and FFs in terms of the dynamical ones. 
Those relations are expected to play 
a critical role to guarantee gauge invariance and the Lorentz invariance of the twist-3 cross sections
to which those DFs and FFs contribute.   Those relations need to be satisfied for a phenomenological
analyses.

\section*{Acknowledgments}

This work has been supported by the Grant-in-Aid for
Scientific Research from the Japanese Society of Promotion of Science
under Contract Nos.~19K03843 (Y.K.) and 18J11148 (K.Y.),
National Natural Science Foundation of China under Project No. 11435004
and research startup funding at South China
Normal University.

\end{document}